\documentclass{amsproc}
\usepackage{oldgerm}
\usepackage{latexsym}
\usepackage{amsmath}
\usepackage{amssymb}
\usepackage{amsgen}
\usepackage{amscd} 
\usepackage{color}
\usepackage{epsfig}

\newtheorem{thm}{Theorem}[section]
\newtheorem{prop}[thm]{Proposition}
\newtheorem{cor}[thm]{Corollary}
\newtheorem{lem}[thm]{Lemma}

\theoremstyle{definition}
\newtheorem{dfn}[thm]{Definition}

\theoremstyle{remark}
\newtheorem{rem}[thm]{Remark}
\theoremstyle{remark}

\newcommand{\opr}[1]{\operatorname{#1}}

\title[A Theory of Physical Quantum Computation]{A Theory of Physical
Quantum Computation: \\
The Quantum Computer Condition} \author[Gerald Gilbert, Michael Hamrick and
F. Javier Thayer]{Gerald Gilbert, Michael Hamrick and
F. Javier Thayer \\
\small \it Quantum Information Science Group$^\ast$$^\dag$\\
{\sc  Mitre} \\
\small \it 260 Industrial Way West, Eatontown, NJ 07724 USA
\footnote{$^\ast$Research supported under MITRE Technology Program Grant 51MSR211.
}
\footnote{$^\dag$E-mail address: \tt{\{ggilbert, mhamrick, jt\}@mitre.org}}
}

\makeindex 

\begin{document} 

\begin{abstract}
In this paper we present a new
unified theoretical framework that describes the full dynamics
of quantum computation.
Our formulation allows any questions pertaining to the physical behavior of
a quantum
computer to be framed, and in principle, answered.
We refer to the central organizing principle developed in this paper,
on which our theoretical structure is based, as the {\em Quantum Computer Condition} (QCC),
a rigorous mathematical statement that connects the irreversible
dynamics of the quantum computing machine, with the reversible operations that comprise the
quantum computation intended to be carried out by the quantum computing machine.
Armed with the QCC, we derive a powerful result that we call
the {\em Encoding No-Go Theorem}. This
theorem gives a precise mathematical statement of the conditions under which
fault-tolerant quantum computation becomes impossible in the presence of dissipation and/or
decoherence. In connection with this theorem, we
explicitly calculate a {\em universal} critical damping value
for fault-tolerant quantum computation.
In addition we show that the recently-discovered approach to quantum
error correction known as ``operator quantum error-correction" is a special case
of our more general formulation. Our approach furnishes what we will refer to as
``operator quantum fault-tolerance." In particular, we show how the QCC allows one to
derive error thresholds for fault tolerance in a completely general context.
We prove the existence of solutions to a class of time-dependent generalizations of
the Lindblad equation.
Using the QCC, we also show that the seemingly
different circuit, graph- (including cluster-) state,
and adiabatic paradigms for quantum computing
are in fact all manifestations of a single, universal
paradigm for all physical quantum computation.
\end{abstract}

\maketitle

\newpage

\tableofcontents

\section{Introduction}

The promise inherent in quantum computing has stimulated a tremendous explosion
of interest in the research community.
Voluminous research has been carried out directed to many different problems associated
with the development and properties of quantum computational algorithms.
Parallel to these efforts, substantial investigations have been devoted to the
problems associated with the development of actual quantum computing machines.
A rigorous and fully general theory that connects quantum computing algorithms
and quantum computing machines would be of considerable value.

In this paper we present a new
unified theoretical framework that describes the full dynamics
of quantum computation.
Our formulation allows any questions pertaining to the physical behavior of
a quantum computer to be framed, and in principle, answered.
We refer to the central organizing principle developed in this paper,
on which our theoretical structure is based, as the {\em Quantum Computer Condition} (QCC),
a rigorous mathematical statement that connects the irreversible
dynamics of the quantum computing machine, with the reversible operations that comprise the
quantum computation intended to be carried out by the quantum computing machine.

The actual dynamics of the
system that we intend to use as a practical quantum computing machine are those of an open
quantum mechanical system, burdened with various dissipative and/or decoherence effects.
The QCC provides a set of mathematical constraints that must be satisfied by a physical
system if
we intend to use that system as a quantum computing machine.

Armed with the QCC, we derive a powerful result that we call
the {\em Encoding No-Go Theorem}. The
Encoding No-Go theorem gives a precise mathematical statement of the conditions under which
fault-tolerant quantum computation becomes impossible in the presence of dissipation and/or
decoherence. We provide a rigorous definition of damping, which includes the phenomena
of dissipation and decoherence, and explicitly
calculate a {\em universal} critical damping value
for fault-tolerant quantum computation.
This fundamental theorem has deep formal significance. Moreover,
it also furnishes criteria for solving diverse problems associated to
actual physical quantum computer realizations, such as
determining which practical design choices for 
quantum computing machines are not viable.

In addition we show that the recently-discovered approach to quantum
error correction known as ``operator quantum error-correction" (OQEC) is
actually a special case
of our general formulation. Our approach furnishes what we will refer to as
``operator quantum fault-tolerance" (OQFT). In
particular, we show how the QCC allows one to
derive error thresholds for fault tolerance in a completely general context.

In this paper we define the concept of a {\em quantum component}, which allows us
to study realistic implementations of quantum computers,
in which decoherence and/or
dissipative effects are present, using a dynamical equation of motion suitable for
describing an open quantum system.
By using the QCC, we are able to reconcile the apparent contradiction between: (1) the
fact that quantum computations are specified by unitary transformations, the
associated dynamics of which are intrinsically reversible, and (2) the fact
that quantum computers,
{\em qua} practical machines, are inevitably characterized by irreversible
dynamics. The reconciliation suggests an analogy with
the {\em fluctuation-dissipation theorem}, which
relates irreversible dynamics to equilibrium properties in a large class of physical
systems.

In this paper we present an existence proof for fundamental solutions to useful
classes of {\em time-dependent} generalizations of the Lindblad equation. This provides
a useful tool in analyzing a wide variety of open quantum mechanical systems.

Our framework is sufficiently general to encompass,
and describe in a unified manner, the currently-known ``paradigms" for quantum computation,
including the {\em circuit-based (``two-way computing") paradigm},
the {\em graph state-based (``one-way computing") paradigm} and
the {\em adiabatic quantum computer paradigm}.
Using the QCC, we show by explicit construction that these seemingly different paradigms
are in fact all manifestations of a single, universal
paradigm for all physical quantum computation.\footnote{
In the particular case of the graph state-based paradigm (which includes cluster
state-based models), we not only show that the
paradigm is a manifestation of the unifying picture provided by the QCC,
but also introduce a definition of graph state-based quantum computers that
generalizes the graph state models previously defined in the literature.
}

\section{The Quantum Computer Condition}\label{QCC_section}

\subsection{Introduction} 

In this section we present the {\em Quantum Computer Condition}, a rigorous mathematical
statement of the constraints that determine the viability of any
practical quantum computing machine.
To achieve the goal of practical quantum computation we must produce an actual
physical device that implements a predetermined unitary operator $U$
acting on some Hilbert space.
The Quantum Computer Condition relates the unitary operator representing a quantum
computation to the actual physical device intended to perform that computation.

The specification of $U$
defines ideally the quantum computation to be performed by the quantum computing
machine.
Generically, 
the result of a quantum computation, $U$, is then used to carry out the
probabilistic evaluation of some classical function.
The complete quantum computation comprises a number of elements, including
\begin{itemize}
\item Preparation of a quantum state for initialization.
\item  Measurement of a quantum state for readout.
\item Various tasks that can be performed by classical computers, such
as preprocessing of the data, or postprocessing of the output into
some humanly comprehensible form.
\end{itemize}
However, the above list of elements are not what is ``important" about quantum
computers. Rather:
\begin{itemize}
\item The distinctive element of quantum computation is the
``ability to perform quantum gates''(c.f.~\cite{nielsen-chuang},
\S4.6). 
\end{itemize}

Mathematically, a quantum gate is a unitary (hence
reversible) operator $U$ acting on a Hilbert space. The formally defined ``gates,"
as such, are not ``devices." They are concepts: they don't implement themselves.
A machine is required to physically
implement the abstractly defined unitary transformation.
The actual, physical computing device intended to implement the transformation
is described mathematically by a completely
positive trace-preserving map, $P$, that transforms the input state to the output state.
We will refer to a physically realizable device intended to implement
an ideal quantum computation as a {\em quantum component}.
In this paper we study
realistic quantum components, in which decoherence and/or
dissipative effects are present, using a dynamical equation of motion suitable for
describing an open quantum system.\footnote{
In order to analyze the effects of dissipation and/or decoherence one must use
{\em some} method of approximating the dynamics of the degrees-of-freedom comprising
the rest of the universe ``outside of" the quantum computer.
This is of course because the complete, detailed,
exact analytical solution to the Schr\"odinger equation of the universe, for all
degrees-of-freedom, is not known.
One reasonable approach is to construct a
Lindblad-type equation, based on a presumption of underlying Markovian dynamics,
in which environment degrees-of-freedom are traced over in such
a way as to result in a first-order (in time) differential equation. In this paper,
for definiteness, we
utilize a generalized
Lindblad-type equation to describe the environment: this is used merely
in order to exemplify
how one may take into account the effects of dissipation and/or decoherence. However,
most of
the results in our paper,
including the crucial Encoding No-Go Theorem,
are independent of this choice, and in particular
are independent of the assumption of underlying Markovian dynamics.}
We must reconcile the fact, and apparent paradox, that a
non-reversible mapping, $P$, is used to ``implement'' a reversible one, $U$.

\subsection{The Motivation of the Quantum Computer Condition}
Mathematically, a quantum computation is a unitary operator $U$ in the
unitary group of a Hilbert space.
A quantum component is described by a
completely positive trace-preserving map $P$ which maps the set of trace class
operators on the Hilbert space to itself. The map $P$ accounts for decoherence and
dissipation,
as well as unitary evolution. We will subsequently discuss in more detail the
actual form for $P$. In our analysis
we will consider the action of $P$ on density matrices
$\rho\mapsto P\cdot\rho$ rather than on state vectors
(and correspondingly the action of $U$ on density matrices $\rho
\mapsto U \rho U^\dag$, rather than the action of $U$ on state
vectors).
This is because, due to the presence of decoherence and/or
dissipation, our system will almost always evolve into a mixed state, which can only
be described by a density matrix $\rho$.\footnote{
Other reasons for utilizing density matrices rather than state vectors include
the generic importance in quantum information theory of trace-preserving
completely positive maps (which restrict to transformations on density
matrices), and the useful algebraic and analytic properties of density matrices
(density matrices for instance form a weak-$\ast$ compact convex set).}

In order to motivate the Quantum Computer Condition (QCC), let us first consider the
abstractly-defined quantum computation itself, prescribed by the unitary operator $U$.
This is assumed to be given, and
is represented by
\begin{equation}
\mathrm{abstract ~computation\!:}\phantom{x} U\rho U^\dag ~.
\end{equation}
The action of the quantum component
intended to effect the computation is represented by
\begin{equation}
\mathrm{practical ~implementation\!:}\phantom{x} P\cdot\rho ~.
\end{equation}
Motivated by the  notion of {\em having the machine implement the computation},
if we were to require that the identity
\begin{equation}\label{ersatz-quantum-computer-equation}
P\cdot\rho = U \rho U^\dag
\end{equation}
hold for all density states $\rho$, then the action of $P$
would in fact be {\em identical} to the action of the unitary operator.
This would imply that $P$ preserves von Neumann entropy~(\cite{vonNeumann},
\S5.3), and hence actually models a system with neither
decoherence nor dissipation, which is not the case for a practical quantum
computing machine. Thus, equation (\ref{ersatz-quantum-computer-equation}) cannot
furnish the correct constraints for realistic quantum computation. We will accordingly
refer to
equation (\ref{ersatz-quantum-computer-equation}) as the {\em ersatz} quantum
computer condition ($\mathcal {E}$QCC).\footnote{
Although the $\mathcal {E}$QCC does not describe practical quantum computing machines,
we note that it can be shown that in the finite-dimensional case, the set of $\rho$
which satisfy~\eqref{ersatz-quantum-computer-equation} is an
algebra $\mathfrak{A}_{P,U}$ which depends on both $P$ and $U$.
The details of how one explicitly
obtains the algebra $\mathfrak{A}_{P,U}$ of solutions to
 ~\eqref{ersatz-quantum-computer-equation}
are given in Appendix~\ref{alpha_0_solution}.}

More realistically, taking into account the inevitable presence of decoherence,
we can require
that~\eqref{ersatz-quantum-computer-equation} hold for some
restricted set of density states. In this case, the solution set will 
correspond to decoherence-free subspaces.
In order to analyze this, we must carefully distinguish between the two different Hilbert
spaces that arise in this problem. The abstract quantum computation is defined
on the Hilbert space of {\em logical} quantum
states, $H_{\mathrm{logical}}$, so that we have 
\begin{equation}
U: H_{\mathrm{logical}}\rightarrow H_{\mathrm{logical}}.
\end{equation}
In contrast, the presence of decoherence (which affects the actual device)
necessitates that
the completely positive trace-preserving map $P$ (which represents the actual device)
is associated to a different Hilbert space,
$H_{\mathrm{comp}}$, the states of which are referred to as
{\em computational} quantum states. The decoherence-free subspace is contained
within $H_{\mathrm{comp}}$.
(The specific decoherence is accounted for in the explicit form of $P$). As
noted above, a consequence
of the decoherence is that $P$ operates on density matrices rather than on
state vectors. Letting ${\mathbf T}(H_{\mathrm{comp}})$ be the Banach space of
trace-class operators on $H_{\mathrm{comp}}$, we have
\begin{equation}
P: {\mathbf T}(H_{\mathrm{comp}})\rightarrow {\mathbf T}(H_{\mathrm{comp}}).
\end{equation}
In order to replace $\mathcal {E}$QCC
(equation (\ref{ersatz-quantum-computer-equation}))
with an equation that properly
incorporates decoherence effects so that it can be used to
determine decoherence-free subspaces, we must introduce suitable encoding and decoding
maps that connect the relevant Hilbert spaces.
We can hope to find an encoding operator
$\mathcal{M}_\mathrm{enc}$ defined on a space of logical inputs and a
decoding operator $\mathcal{M}_\mathrm{dec}$ with values in a space of
logical outputs, the actions of which are given by
(here ${\mathbf T}(H_{\mathrm{logical}})$ is the Banach space of trace-class operators
on $H_{\mathrm{logical}}$, analogous to ${\mathbf T}(H_{\mathrm{comp}})$)
\begin{equation}
\mathcal M_{\mathrm{enc}}: {\mathbf T}(H_{\mathrm{logical}}) \rightarrow
{\mathbf T}(H_{\mathrm{comp}})
\end{equation}
and
\begin{equation}
\mathcal M_{\mathrm{dec}}: {\mathbf T}(H_{\mathrm{comp}}) \rightarrow
{\mathbf T}(H_{\mathrm{logical}}),
\end{equation}
such that ({\em cf} equation (\ref{ersatz-quantum-computer-equation}))
\begin{equation}\label{encoded-quantum-computer-equation}
\mathcal{M}_\mathrm{dec}(P\cdot(\mathcal{M}_\mathrm{enc}(\rho))) =
U\rho U^\dag
\end{equation}
for all logical inputs $\rho$. We will refer to equation
(\ref{encoded-quantum-computer-equation}) as the ``encoded {\em ersatz}
quantum computer condition (e$\mathcal {E}$QCC).\footnote{
Note that no encoding map is required on the right-hand side of equation
(\ref{encoded-quantum-computer-equation}) since the unitary $U$ by
definition acts on $H_{\mathrm{logical}}$.} The existence of the encoding and decoding maps
$\mathcal M_{\mathrm{enc}}$ and $\mathcal M_{\mathrm{dec}}$ is a consequence of
the presumed existence of an associated decoherence-free subspace of $H_{\mathrm{comp}}$,
of dimension greater than or equal to the dimension of $H_{\mathrm{logical}}$.

The meaning of e$\mathcal {E}$QCC given in eq.(\ref{encoded-quantum-computer-equation})
is as follows. Given a chosen quantum computation, $U$, we wish to construct a
physical ``machine," $P$, that implements $U$. In order to do this we must
find 
encoding and decoding maps $\mathcal{M}_\mathrm{enc}$ and $\mathcal{M}_\mathrm{dec}$
such that the equation is satisfied for all $\rho$. This is a crucial difference
between eqs.(\ref{encoded-quantum-computer-equation}) and
(\ref{ersatz-quantum-computer-equation}):
requiring that eq.(\ref{ersatz-quantum-computer-equation}) holds for
all $\rho$ implies a machine that preserves von Neumann
entropy, does not dissipate heat and does not decohere,
and thus does not describe a practical quantum computing device. In contrast,
eq.(\ref{encoded-quantum-computer-equation}) holds for all states $\rho$, but does
so by making use of the encoding and
decoding operators to map to a decoherence-free subspace.
The e$\mathcal {E}$QCC (eq.(\ref{encoded-quantum-computer-equation}))
formally holds for all
density states $\rho$, similar to eq.(\ref{ersatz-quantum-computer-equation}),
but the use of the encoding and decoding maps
in eq.(\ref{encoded-quantum-computer-equation}) effectively confines the solution
to a restricted set in $H_{\mathrm{comp}}$.

However, eq.(\ref{encoded-quantum-computer-equation}) does not in general provide
an acceptable condition to connect the dynamics of a practical quantum computing
device to the constraints implied by the unitary operator $U$ that defines the
abstract quantum computation. {\em {This is because the formulation presented by
\eqref{encoded-quantum-computer-equation} does not address
situations in which residual errors cannot
be completely eliminated}},
even with the use of decoherence-free subspaces,
and/or other error correction methods~\cite{lbw_dfs_qecc_99}, \cite{shor_ft_96}.

To recapitulate, eq.(\ref{ersatz-quantum-computer-equation}) describes a quantum
computer that performs the required computation $U$, but only if the implementing device,
described by the completely positive map $P$, neither dissipates heat nor decoheres. We thus
reject this expression as a viable quantum computer condition
because it describes a system that is effectively impossible to
achieve. In contrast, eq.(\ref{encoded-quantum-computer-equation})
describes a quantum computer that performs the required computation $U$,
but only if the device described by the completely positive map $P$ implements
the required decoherence-free subspace with absolutely no residual errors.
This does not provide a sufficiently
general formulation: {\em we need to consider situations in which residual errors cannot
be completely eliminated}.

\subsection{The Presentation of the Quantum Computer Condition}
We wish to allow for the likelihood that, even if a decoherence-free subspace
can be found, and even if error correction procedures are applied, there will still
be residual errors characterizing the operation of the
quantum computer. Such a situation may arise even if
error correction is correctly applied: for instance, in the application of concatenated
error codes, one iterates the concatenation process until the error probability is
reduced to a value that is deemed ``acceptable"~\cite{shor_ft_96}, \cite{preskill_ft_97}.
This final error
probability, though small, is not exactly zero. The important point is that
it is prudent to write down our
quantum computer condition so as to reflect the
inevitable survival of some amount of residual error.

In order to quantify the degree to which the actual computational device,
represented by $P$, {\em cannot exactly} (because of residual error) implement the
ideal quantum computation, represented by $U$, we consider the following difference
({\em cf} eq.\eqref{encoded-quantum-computer-equation})
\begin{equation}\label{inaccuracy_preparameter}
\mathcal{M}_\mathrm{dec}(P\cdot(\mathcal{M}_\mathrm{enc}(\rho))) -
U\rho U^\dag.
\end{equation}
We now compute for this difference a suitable norm on matrices (this norm is made more
precise below), as
\begin{equation}\label{inaccuracy_parameter}
\|\mathcal{M}_\mathrm{dec}(P\cdot(\mathcal{M}_\mathrm{enc}(\rho))) -
U\rho U^\dag\|.
\end{equation}
This quantity is of fundamental importance: {\em it
is a measure of the inaccuracy of the implementation of $U$
by} $P$. It tells us how well a practical quantum computing
device actually implements an ideally-defined quantum computation. We will refer
to the scalar quantity
given by (\ref{inaccuracy_parameter}) as the {\em implementation inaccuracy}
associated to the pair $U$ and $P$.
In connection with this, we introduce a parameter, $\alpha$, to specify the maximum
tolerable implementation inaccuracy, so that we have
\begin{equation}\label{pre_encoded_QCC}
\|\mathcal{M}_\mathrm{dec}(P\cdot(\mathcal{M}_\mathrm{enc}(\rho))) -
U\rho U^\dag\| \leq \alpha .
\end{equation}

\subsubsection{Formal Statement of the Quantum Computer Condition (QCC)}
Motivated by the above considerations,
we now introduce and formally define an inequality of fundamental importance in the theory of
quantum computation that
we will refer to as the {\em Quantum
Computer Condition}. We will formally designate this condition
by $\mathbf{QCC}(P,U,\mathcal{M}_\mathrm{enc},
\mathcal{M}_\mathrm{dec},\alpha)$.
In the following, we will impose no constraint on the dimensionality of the Hilbert
space, and in particular we allow Hilbert spaces of infinite dimensions.
Let $U$ be a unitary on
$H_{\mathrm{logical}}$ and
$P$ be a trace-preserving completely positive map on
$\mathbf{T}(H_\mathrm{comp})$. 
Let
$\mathcal M_{\mathrm{enc}}: {\mathbf T}(H_{\mathrm{logical}}) \rightarrow
{\mathbf T}(H_{\mathrm{comp}})$ and
$\mathcal M_{\mathrm{dec}}: {\mathbf T}(H_{\mathrm{comp}}) \rightarrow
{\mathbf T}(H_{\mathrm{logical}})$ be
completely-positive, trace-preserving encoding and decoding maps
(with no further restrictions of any kind on
the encoding and decoding maps).
The Quantum Computer Condition,
$\mathbf{QCC}(P,U,\mathcal{M}_\mathrm{enc},
\mathcal{M}_\mathrm{dec},\alpha)$, holds iff
for all density matrices $\rho \in \mathbf{T}(H_\mathrm{logical})$, we have
\begin{equation}\label{encoded_QCC}
\|\mathcal{M}_\mathrm{dec}(P\cdot(\mathcal{M}_\mathrm{enc}(\rho))) -
U\rho U^\dag\|_1 \leq \alpha ,
\end{equation}
where $\mathbf{T}(H_\mathrm{comp})$ and $\mathbf{T}(H_\mathrm{logical})$ are
the Banach spaces of trace-class operators on $H_\mathrm{comp}$ and
$H_\mathrm{logical}$, respectively, and
$\|\cdot\|_1$ is the Schatten $1$-norm defined in
Appendix~\ref{banach-spaces}.

It should be noted that an alternate measure of distance between density matrices
to that provided by the Schatten $1$-norm is given by the fidelity
function \cite{nielsen-chuang}.
One could write an alternate form of the QCC in terms of the fidelity that would be
essentially equivalent to the form of the QCC given in \eqref{encoded_QCC} above.
Using an obvious notation to denote the QCC written with each of these definitions of
distance, the two forms are related as follows. Given a quartet
$\{ P,U,\mathcal{M}_\mathrm{enc},
\mathcal{M}_\mathrm{dec}\}$, one can show that if
$\mathbf{QCC}_{{\mathrm{Schatten}}}(P,U,\mathcal{M}_\mathrm{enc},
\mathcal{M}_\mathrm{dec},\alpha)$ is satisfied, then the fidelity-based version
of QCC given by
$\mathbf{QCC}_{{\mathrm{fidelity}}}(P,U,\mathcal{M}_\mathrm{enc},
\mathcal{M}_\mathrm{dec},\alpha^\prime)$ is also satisfied, where
$\alpha^\prime = \alpha^\prime(\alpha)$ is a function of $\alpha$.
The form of QCC based on the Schatten $1$-norm given above in \eqref{encoded_QCC}
is more convenient for our purposes for a number of 
mathematical reasons, including the fact that that the fidelity, as such, is not
a proper norm. For instance, the
statement and proof of the Encoding No-Go Theorem carried out in
\S\ref{NGT} below are more conveniently presented making use of the form of the QCC
based on the Schatten norm.

Note that $\mathcal{M}_\mathrm{enc}$ and $\mathcal{M}_\mathrm{dec}$
do not 
represent {\em physical} operations: all physical operations are
carried out by the quantum component $P$.
If the proper distinction between these maps and $P$ 
is not observed, one could include the {\em entire} computation in the definition of the
maps, with 
the absurd conclusion that any quantum computation could be performed in the absence of any 
real hardware.

\subsubsection{Some implications of the QCC}

The QCC is a remarkably powerful expression. It constitutes a kind of ``master
expression" for physical quantum computation. The inequality \eqref{encoded_QCC}
concisely incorporates a complete specification of the full dissipative, decohering
dynamics of the actual,
practical device used as the quantum computing machine, a specification of
the ideally-defined quantum
computation intended to be performed by the machine, and a quantitative criterion
for the accuracy with which the computation must be executed given the inevitability
of residual errors surviving even after error correction has been applied.

Making use of the QCC, one can state and
prove (we do this is in \S\ref{NGT}, the next section of the paper)
a fundamental and powerful theorem in the subject of quantum computing,
the {\em Encoding No-Go Theorem}. This
no-go theorem gives a precise mathematical statement of the conditions under which
fault-tolerant quantum computation becomes impossible in the presence of dissipation and/or
decoherence. Apart from its formal significance, the theorem can be used to compare
different proposed physical approaches to actually building a quantum computing
machine, with the no-go condition furnishing a criterion for the practical
engineering viability of various choices.

As a further indication of the power and general applicability of the QCC, we
show that one can apply it
to the known, seemingly distinct ``paradigms" for quantum computing, based on (1) the
use of quantum circuits built up out of quantum gates (the {\em circuit-based paradigm},
or ``two-way" quantum computing), (2) the use of graph states or cluster states
(the {\em graph state-based paradigm}, or ``one-way" quantum computing) and (3) the
use of specially chosen Hamiltonians describing adiabatic dynamics (the {\em adiabatic
quantum computer paradigm}). The QCC allows one to show that these apparently
different definitions of a quantum computer are in fact manifestations of the same
underlying formulation: {\em there is only one paradigm for quantum computers}.
The application of the QCC to different quantum computing paradigms
is discussed in \S\ref{unification}.

The encoding-decoding pair $\mathcal{M}_\mathrm{enc}$ and $\mathcal{M}_\mathrm{dec}$
that appear in the QCC are defined quite generally as completely positive
trace-preserving maps. This formulation is sufficiently general to encompass
all possible encodings associated with standard quantum error correction (QECC)
techniques, decoherence-free subspaces and noiseless subsystems. More generally still,
we show below in \S\ref{KLP_reduction} that the recently discovered approach
known as ``operator quantum error correction" (OQEC) is in fact a special case
of our more general QCC formulation.
In addition the QCC can be used to extend OQEC to
what we will refer to as
``operator quantum fault-tolerance" (OQFT). In particular, we show in \S\ref{et_ft} below
how the QCC allows one to
derive error thresholds for fault tolerance in a completely general context.

Another significant consequence of the QCC is that it resolves the apparent
paradox that the
quantum computations we wish to perform are defined by reversible operators, but
the actual devices that we must use to execute the computations are
necessarily described by irreversible maps.
We note that this is reminiscent of the {\em fluctuation-dissipation theorem}, which
relates irreversible dynamics to equilibrium properties in a large class of physical
systems. Inspection of \eqref{encoded_QCC} reveals
that the paradox is resolved through the transformations provided by the
encoding and decoding maps associated to the QCC. Roughly speaking,
reversible behavior of the actual device is enabled only on the code subspace
defined by $\mathcal{M}_\mathrm{enc}$ and $\mathcal{M}_\mathrm{dec}$.

Note that if one describes quantum computing solely in terms of the
unitary transformations that define the computations,
it is not too surprising that the resulting computational model
is in some way ``powerful."
After all, unitary transformations on finite dimensional
spaces include such powerful operations as the discrete Fourier
transform which are known to play an important role in number
theoretic problems.
Rather than regarding the power of the ideally-defined quantum computation as the
remarkable thing, the {\em truly} remarkable thing would be the
construction of an inherently irreversible
device that actually implements the reversible, unitary map to a
specified level of accuracy. It is this possibility that our QCC expresses
and makes mathematically precise. The QCC can thus be used to formulate and prove
assertions about the physical solvability of particular computational
problems.

\subsubsection{Operator quantum error correction (OQEC) as a special case
of QCC}\label{KLP_reduction}

The recently developed theory of operator quantum error correction
\cite{klp_05}, \cite{klpl_05}
unifies many apparently disparate approaches to the practical 
problem of dealing with errors in quantum information.  Among these approaches are 
quantum error correction, decoherence 
free subspaces and noiseless subsystems.
Here we show that OQEC is in fact a special case of the general statement of the
QCC. In this section we demonstrate that the entire formalism of OQEC can be obtained
from QCC by
choosing $\mathcal{M}_\mathrm{enc}$ and $\mathcal{M}_\mathrm{dec}$ as described below,
and by setting setting $U=I$ and $\alpha=0$ in the QCC, so that the reduction
QCC $\rightarrow$ OQEC is given by:
\begin{equation}
\mathbf{QCC}(P,U,\mathcal{M}_\mathrm{enc},\mathcal{M}_\mathrm{dec},\alpha)\rightarrow
\mathbf{QCC}(P,I,\mathcal{M}_\mathrm{enc},\mathcal{M}_\mathrm{dec},0)~.
\end{equation}

In the scheme of OQEC, the techniques
of (standard) quantum error correction, decoherence 
free subspaces and noiseless subsystems are subsumed 
under the 
unified concept of ``correctability." This is defined formally as follows.    
Let $H_\mathrm{comp}$ be a Hilbert space with a decomposition 
$H_\mathrm{comp}=\left( H^A \otimes H^B\right) 
\oplus K$, where $H^A$, $H^B$ and $K$ are discussed in the following paragraph.
We identify this as the computational space $H_\mathrm{comp}$ 
of this paper because, as we shall see below, it describes the space on which the
real physical processes of error and recovery operate.
Let $\mathfrak{S}$ be the semigroup given by 
\begin{equation}
\mathfrak{S} = \{ \sigma \in \mathbf{L}(H_\mathrm{comp}) : 
   \exists \sigma^A \in \mathbf{L}(H^A), \exists \sigma^B \in \mathbf{L}(H^B), ~
   \mathrm{such~that} ~ \sigma = \sigma^A \otimes \sigma^B \}~,
\end{equation}
where $\mathbf{L}(\cdot)$ is the space of bounded operators\footnote{In \S\ref
{KLP_reduction} of this paper we assume (following \cite{klpl_05}) that
all Hilbert spaces are finite-dimensional.
Although not discussed in \cite{klpl_05},
it is important to note that if one makes this assumption,
there is then
no need to distinguish between bounded operators in $L(\cdot )$
and trace-class operators in $T(\cdot )$.
This distinction is important in the other sections of our paper
where we allow Hilbert spaces of infinite as well as finite dimensionality.} on
the appropriate Hilbert
space, with the operator norm,
and let $\mathcal{E}$ and $\mathcal{R}$ be completely-positive, trace preserving 
maps on $\mathbf{L}(H_\mathrm{comp})$ corresponding to error processes and 
recovery processes, 
respectively.  We say that $\mathfrak{S}$ is correctable for $\mathcal{E}$ iff
\begin{equation}\label{klpcorrectable}
\forall \sigma \in \mathfrak{S},~ 
\left( \mathrm{Tr}_A \circ \mathcal{P}_\mathfrak{S} \circ \mathcal{R} 
   \circ \mathcal{E} \right) \sigma = \mathrm{Tr}_A \sigma ~,
\end{equation}
where $\mathcal{P}_\mathfrak{S}$ effectively projects density matrices onto the
$ H^A \otimes H^B$ subspace of $H_{\mathrm{comp}}$.

Physically speaking, $H^B$ is the Hilbert space which carries the information to 
be protected from errors, the ``noiseless subsystem," 
while $H^A$ is the Hilbert space on which the errors are 
permitted to operate freely, the ``noisy subsystem."  
$K$ is the orthogonal complement of $H^A \otimes H^B$ in $H_\mathrm{comp}$ 
and is simply projected out by $\mathcal{P}_\mathfrak{S}$.
Thus, in \eqref{klpcorrectable}, 
$\mathrm{Tr}_A \sigma$ represents the quantum information which is to be protected.  
The left side of \eqref{klpcorrectable} describes the effect of allowing  
errors to operate on the full state $\sigma$, and then applying 
recovery procedures.  (The projection and the 
trace simply extract the state of the noiseless subsystem.)  According to 
\eqref{klpcorrectable}, correctability thus means that 
the recovery procedure does in fact recover the state of the noiseless subsystem, 
$\mathrm{Tr}_A \sigma$, without error.  

As we now show, the definition of correctability is actually a special case of the QCC.  
Note first that correctability, as 
defined above, applies to a quantum channel as opposed to a quantum computer.  
It describes the transportation of a quantum state in a noisy environment as opposed to the 
``processing" of a quantum state so as to implement a quantum computation.  
In order to make the connection with the QCC, we may 
therefore think of the quantum channel as a quantum computer that implements the 
identity operation:
\begin{equation}
U = I_{H_{\mathrm {logical}}}
\end{equation}

In the definition of correctability, the part of the state containing the 
information of interest is 
recovered without error.  This corresponds to taking 
$\alpha = 0$ in the QCC.  Note that, as discussed above, this is not practically 
achievable for quantum computers.  
This is less obviously an issue for the theory of operator error correction as currently 
formulated \cite{klp_05}, \cite{klpl_05}, since that theory addresses a relatively
circumscribed set of circumstances in which one is concerned with transporting quantum 
states rather than implementing a quantum computation.  In particular, the error process 
$\mathcal{E}$, which is specified in advance, only operates {\em once} on the quantum state 
being transported. Error processes associated 
with the constituent parts of quantum computers operate each time the constituent part 
operates on the 
the quantum state being processed. The problem of fault 
tolerant quantum computation is inherently more complex than the problem of 
error correction/protection for a quantum channel.  We will
discuss this more fully in \S\ref{et_ft} below.

Setting $U=I$ and $\alpha = 0$ in the quantum computer condition, we obtain
\begin{equation}
\|\left( \mathcal{M}_\mathrm{dec}\circ P \circ \mathcal{M}_\mathrm{enc} \right)\rho -
\rho\|_1 = 0 ,
\end{equation}
or
\begin{equation}
\left (\mathcal{M}_\mathrm{dec} \circ P \circ \mathcal{M}_\mathrm{enc}
\right) \rho = \rho ~,
\end{equation}
where $\rho\in L(H_{{\mathrm{logical}}})$.\footnote{As
noted in the previous footnote, we are assuming in \S\ref{KLP_reduction}
that $H_{{\mathrm{logical}}}$ is finite-dimensional.}

We now proceed to define the encoding map $\mathcal{M}_\mathrm{enc}$. For this purpose
we define a map 
$W_\mathrm{enc}:\mathbf{L}(H_\mathrm{logical}) \rightarrow \mathbf{L}(H^B)$
that encodes the logical quantum state $\rho$ in a state $\sigma^B$ of the noiseless subsytem.  We 
further define a map $W_\mathrm{adj}(\sigma^A):\mathbf{L}(H^B) 
\rightarrow \mathbf{L}(H^A \otimes H^B)$ 
that adjoins an arbitrary state $\sigma^A$ of the noisy subsystem to the state $\sigma^B$, 
that is 
\begin{equation}
W_\mathrm{adj}(\sigma^A) : \sigma^B \mapsto \sigma \equiv \sigma^A \otimes \sigma^B
\end{equation}
We then define the full encoding map that appears in the QCC as follows:
\begin{equation}
\mathcal{M}_\mathrm{enc} \equiv W_\mathrm{adj}(\sigma^A) \circ W_\mathrm{enc}
\end{equation}

The map $P$ characterizes the dynamics of the physical computer, which in this case is just 
the (noisy) channel followed by the recovery procedure:
\begin{equation}
P \equiv \mathcal{R} \circ \mathcal{E}
\end{equation}
We note that the current formulation of the theory of operator quantum error correction 
implicitly assumes that the recovery process $\mathcal{R}$ can be implemented without error, 
even though this requires, in general, that coherent operations be performed on entangled 
quantum states.  Once again, 
the emphasis on error correction alone avoids the more difficult issue of 
achieving true fault tolerance.

Finally we define the decoding map as:
\begin{equation}
\mathcal{M}_\mathrm{dec} = W_\mathrm{enc}^{-1} \circ \mathrm{Tr}_A \circ \mathcal{P}_\mathfrak{S}~,
\end{equation}
which extracts the state of the noiseless subsystem and decodes it to obtain a state in the 
logical space $\mathbf{L}(H_\mathrm{logical})$.  

With the above definitions, the QCC becomes 
\begin{equation}
\left( W_\mathrm{enc}^{-1} \circ \mathrm{Tr}_A \circ \mathcal{P}_\mathfrak{S} \circ
       \mathcal{R} \circ \mathcal{E} \circ W_\mathrm{adj}(\sigma^A) \circ W_\mathrm{enc} \right)
\rho = \rho~.
\end{equation}
Applying $W_\mathrm{enc}$ to both sides, and recalling that $W_\mathrm{enc} \rho = 
\sigma^B = \mathrm{Tr}_A \sigma$, we obtain
\begin{equation}
\left( \mathrm{Tr}_A \circ \mathcal{P}_\mathfrak{S} \circ
       \mathcal{R} \circ \mathcal{E} \circ W_\mathrm{adj}(\sigma^A) \circ W_\mathrm{enc} \right)
\rho = \mathrm{Tr}_A \sigma~.
\end{equation}
Noting that 
$W_\mathrm{adj}(\sigma^A) \circ W_\mathrm{enc} \rho = \sigma$, we obtain the correctability 
condition of~\cite{klp_05}, \cite{klpl_05}:
\begin{equation}
\left( \mathrm{Tr}_A \circ \mathcal{P}_\mathfrak{S} \circ
       \mathcal{R} \circ \mathcal{E} \right)
\sigma = \mathrm{Tr}_A \sigma~.  
\end{equation}

In summary, we have shown that the formalism of operator quantum error correction
actually arises as a special case of an underlying definition of physical quantum
computation given by the QCC. In addition, examination of operator quantum error correction
(OQEC) from the perspective of the QCC reveals limitations and restrictions implicit to the
formalism of operator quantum error correction, and inherited from the previously 
known, standard approaches to error correction (QECC). These limitations
render direct application of either OQEC or QECC to questions of fault tolerance somewhat
problematic, whereas the QCC approach is more immediately applicable.
Thus, the QCC enables one to generalize OQEC to operator fault tolerance (OQFT).

\section{The Encoding No-Go Theorem}\label{NGT}

\subsection{Introduction}
Armed with the QCC, in this section we state and prove a theorem of crucial
importance in the theory of physical quantum computation. This is the {\em Encoding
No-Go Theorem}, which
gives a precise mathematical statement of the conditions under which
fault-tolerant quantum computation becomes impossible in the presence of damping.
Damping, for which we provide a mathematically rigorous definition below,
includes the effects of dissipation and decoherence.
The No-Go theorem for a completely positive trace-preserving
map $P$ corresponding to a putative quantum
computing device then asserts that, in the presence of sufficient damping,
the quantum computer condition
$\mathbf{QCC}(P, U, \mathcal{M}_\mathrm{enc},
\mathcal{M}_\mathrm{dec},\alpha)$ ({\em cf} \eqref{encoded_QCC})
cannot be satisfied for any
encoding-decoding pair, unless $H_\mathrm{logical}$
has dimension $1$: there is then effectively no quantum computer.
(In the case that dim $H_\mathrm{logical}=1$ we are of course unable to define
a meaningful quantum computation at all.)
As part of the No-Go Theorem we explicitly calculate a {\em universal} critical damping
value for fault-tolerant quantum computation.
We precisely state and prove this theorem in the remainder of \S\ref{NGT}.

\subsection{Encoding and Decoding Maps}
An encoding-decoding pair are completely positive, trace-preserving maps
\begin{equation}
\begin{aligned}
& \mathcal{M}_\mathrm{enc}:\mathbf{T}(H_\mathrm{logical}) \rightarrow \mathbf{T}(H_\mathrm{comp})\\
& \mathcal{M}_\mathrm{dec}:\mathbf{T}(H_\mathrm{comp}) \rightarrow \mathbf{T}(H_\mathrm{logical}).
\end{aligned}
\end{equation}
Encoding-decoding pairs provide the link between a completely positive
map $P:\mathbf{T}(H_\mathrm{comp}) \rightarrow
\mathbf{T}(H_\mathrm{comp})$ corresponding to a physical device and a
unitary operator $U:H_\mathrm{logical} \rightarrow H_\mathrm{logical}$
corresponding to a quantum computation. We will
place no further restrictions on encoding-decoding pairs.
Now, suppose $\mathcal{M}_\mathrm{enc}, \mathcal{M}_\mathrm{dec}$ is an
encoding-decoding pair.  
Then the adjoint maps $\mathcal{M}_\mathrm{dec}^\mathrm{t}$,
$\mathcal{M}_\mathrm{enc}^\mathrm{t}$ are unit preserving completely
positive maps. (Adjoint maps of completely positive maps are defined
in \eqref{definition-of-dual} in Appendix \ref{mathematical-preliminaries-section}.)

\subsection{Damping and $\gamma$-damping}
Dissipative and decohering effects are consequences of {\em damping}.
To begin the development of the No-Go Theorem, we introduce
a mathematically precise definition of the damping
of a quantum mechanical system, to which we refer as $\gamma$-damping.

\begin{dfn}\label{gamma-damping}
Let $H$ be a Hilbert space. A completely positive trace-preserving map
$P:\mathbf{T}(H) \rightarrow \mathbf{T}(H)$ is {\em $\gamma$-damped}
iff there is an abelian von Neumann algebra $\mathfrak{A} \subseteq
\mathbf{L}(H)$ such that for all $T \in \mathbf{L}(H)$, there is
an $S_T \in \mathfrak{A}$ such that
\begin{equation}\label{gamping}
\|P^\mathrm{t}(T)- S_T\|_\infty \leq \gamma  \|T\|_\infty,
\end{equation}
\end{dfn}
\noindent where the operator norm $\|\cdot\|_\infty$ is
defined in Appendix~\ref{banach-spaces}.
Note that larger values of $\gamma$ correspond to less damping of the system.

To make contact with the physically intuitive notion of
damping, we apply this definition to the example of the simple
harmonic oscillator subject to phase damping.
For such an harmonic oscillator,
the $ij$th element of the density matrix, $\rho_{ij}=\langle i|\rho
|j\rangle$, decays exponentially as $e^{-\kappa (i-j)^2}$, where we are working
in the basis of energy eigenstates identified by the labels $i$ and $j$.
The quantity $\kappa$ is characteristic of the specific 
oscillator and its coupling to the environment.
The
completely positive map $P$ transforms the initial, general density
matrix for the system into a density matrix with exponentially-decaying
off-diagonal elements. Under the influence of damping, as the
off-diagonal states of the oscillator decay and approach zero, the
density matrix converges to a diagonal density matrix
in the $\{ ij\}$ basis specified above.
This is true for all initial configurations of the oscillator,
and thus the damping process tends to an abelian set of final configurations.
(The damping
parameter $\kappa$ that characterizes the decay of the off-diagonal
elements of the density matrix is related to the quantity $\gamma$
that appears in \eqref{gamping}.
As noted above, larger values of $\gamma$ correspond to less damping and
hence smaller values of $\kappa$.)

\subsection{No-Go Theorem for Encodings}

\noindent{We now state the main result of \S\ref{NGT}:
{\em the Encoding No-Go Theorem}.}
\begin{thm}[\bf {The Encoding No-Go Theorem}]\label{ngt1}~\\
Suppose
that $\mathbf{QCC}(P,U,\mathcal{M}_\mathrm{enc},
\mathcal{M}_\mathrm{dec},\alpha)$ holds. If
$P:\mathbf{T}(H_\mathrm{comp}) \rightarrow
\mathbf{T}(H_\mathrm{comp})$ is $\gamma$-damped and $2\gamma + \alpha < \sqrt{2}/4$,
then $H_\mathrm{logical}$ has dimension $1$.
\end{thm}
\noindent To prove the Encoding No-Go Theorem, we first derive
in \S\ref{ab_factor} a number of general mathematical
results on completely positive maps. We then apply these
specifically to obtain a proof of the Encoding No-Go Theorem in \S\ref{ng_proof} below.

\subsection{Completely Positive Maps with Abelian
Factorizations}\label{ab_factor} 

We begin with the following lemma, which furnishes
a superoperator version of the definition of $\gamma$-damping.
For any abelian von Neumann algebra $\mathfrak{A} \subseteq
\mathbf{L}(H)$ there is a unit-preserving completely positive
projection operator $\mathbf{E}_\mathfrak{A}:\mathbf{L}(H) \rightarrow
\mathfrak{A}$.  This fact is elementary, but also follows from
injectivity~\cite{connes} of such algebras, in the case $H$ is
separable.
\begin{lem}\label{superop_gamma}
If $P$ is $\gamma$-damped, $\mathfrak{A}$ is as given in Definition~\ref{gamma-damping}
and $\mathbf{E}_\mathfrak{A}:\mathbf{L}(H)
\rightarrow \mathfrak{A}$ is a unit-preserving completely positive
projection mapping $\mathbf{E}_\mathfrak{A}:\mathbf{L}(H) \rightarrow
\mathfrak{A}$, then
\begin{equation}
\|P^\mathrm{t}(T) - \mathbf{E}_\mathfrak{A}P^\mathrm{t}(T)\|_\infty \leq 2 \gamma \| T\|_\infty.
\end{equation}
\end{lem}
\begin{proof}
Note that $\mathbf{E}_\mathfrak{A}$ is a linear mapping of norm $\leq
1$ and thus
\begin{equation}
\begin{aligned}
\|P^\mathrm{t}(T) - \mathbf{E}_\mathfrak{A}P^\mathrm{t}(T)\|_\infty & \leq \|P^\mathrm{t}(T) -
S_T\|_\infty + \|S_T - \mathbf{E}_\mathfrak{A}P^\mathrm{t}(T)\|_\infty \\
&  \leq \gamma  \| T\|_\infty + \|\mathbf{E}_\mathfrak{A} S_T - \mathbf{E}_\mathfrak{A}P^\mathrm{t}(T)\|_\infty \\
& \leq 2 \gamma \| T\|_\infty,
\end{aligned}
\end{equation}
as claimed.
\end{proof}

In the remainder of this section we prove an important
technical result used in the proof of the no-go theorem, namely that
completely positive maps $F: \mathbf{L}(H) \rightarrow \mathbf{L}(H)$,
which factor through completely positive maps into abelian von Neumann
algebras, cannot be used to approximate unitary operators $U$ on $H$
if $\opr{dim}(H) \geq 2$. More precisely, we will show that for any
$\beta < \sqrt{2}/4$, there is at least one non-zero operator $T$ for which
\begin{equation}\label{non-approximation-inequality}
\|U^\dagger T U - F(T) \|_\infty \geq \beta \  \| T\|_\infty.
\end{equation}
We first define what it means for a completely positive map to
factor through an abelian von Neumann algebra:
\begin{dfn}\label{defn-of-abelian-factorization}
Let $A$ be a $\mathrm{C}^\ast$-algebra with a multiplicative unit.  A
unit preserving completely positive map $F:A \rightarrow A$ has an
{\em abelian factorization} (or briefly is abelian factorizable) iff
there is an abelian von Neumann algebra $\mathfrak{B}$ and unital
completely positive maps $Q:A \rightarrow \mathfrak{B}$, $R :
\mathfrak{B} \rightarrow A$ such that $F$ is the composition $R \circ
Q$.
\end{dfn}
If $F: \mathbf{L}(H) \rightarrow \mathbf{L}(H)$ is abelian
factorizable, then it follows from the definition that for any
unit-preserving completely positive maps $Q, R$, the completely
positive map $Q \circ F \circ R$ is abelian factorizable.
We begin by providing a characterization of abelian factorizable maps.
\begin{prop}
Let $A$ be an arbitrary $\mathrm{C}^\ast$-algebra with multiplicative
unit.  A unit preserving completely postive map $F: A \rightarrow A$
factors through a finite-dimensional abelian von-Neumann algebra iff
there are unit preserving positive linear functionals $\rho_1, \ldots,
\rho_m \in A^\dagger$ and positive elements $G_1,
\ldots, G_m \in A$ of norm $\leq 1$, such that $\sum_{i=1}^m G_i = I$
and
\begin{equation}\label{finite-factorization-equation}
F(T) = \sum_{i=1}^m \rho_i(T) G_i \quad \forall T \in A.
\end{equation}
\end{prop}
\begin{proof}
We first note that positive maps from an abelian
$\mathrm{C}^\ast$-algebra or into an abelian $\mathrm{C}^\ast$-algebra
are automatically completely positive (see~\cite{stinespring1955},\cite{choi1972}).
Thus the map $F$ given by
Equation~\eqref{finite-factorization-equation} is completely positive.
Let $\mathfrak{B}$, $Q:A \rightarrow \mathfrak{B}$, $R : \mathfrak{B}
\rightarrow A$ as in Definition~\ref{defn-of-abelian-factorization},
but with $\mathfrak{B}$ finite dimensional and let $E_1, \ldots, E_m$
be the minimal non-zero projections of $\mathfrak{B}$.  Then
\begin{equation}
F(T) = R\biggl(\sum_{i=1}^m E_i Q(T) E_i\biggr) = R\biggl(\sum_{i=1}^m
\rho_i(T) E_i\biggr) = \sum_{i=1}^m \rho_i(T) R(E_i).
\end{equation}
Letting $G_i = R(E_i)$,
\begin{equation}
\sum_{i=1}^m G_i = \sum_{i=1}^m R(E_i) = R\biggl(\sum_{i=1}^m E_i\biggr) = I.
\end{equation}
This completes the proof.
\end{proof}
In general, unit-preserving completely positive maps with arbitrary
abelian factorizations can be approximated by maps of the
form~\eqref{finite-factorization-equation}.
\begin{prop}\label{approximation-result-for-abelain-factorizable}
If a completely positive map $F:A \rightarrow A$ has an abelian
factorization, then there is a generalized sequence of maps
$\{F_\kappa\}_{\kappa \in K}$ each having the form $F_\kappa(T) =
\sum_{i=1}^m \rho_i(T) G_i$ which converges to $F$ in the point-norm
topology, that is for each $T \in A$,
$F_\kappa(T) \rightarrow F(T)$ in the norm of $A$.
\end{prop}
\begin{proof}
Let $\mathfrak{B}$, $Q:A \rightarrow \mathfrak{B}$, $R : \mathfrak{B}
\rightarrow A$ as in
Definition~\ref{defn-of-abelian-factorization}. Given $T_1, \ldots,
T_m \in A$, let $\mathfrak{B}_0$ be a finite dimensional abelian
von-Neumann subalgebra of $\mathfrak{B}$ and $\mathbf{E}$ a linear
projection operator\footnote{These operators are sometimes referred to
as {\em conditional expectations}.} $\mathfrak{B} \rightarrow
\mathfrak{B}_0$ such that \begin{equation}
\|Q(T_i) - \mathbf{E}(Q(T_i))\|_\infty \leq \epsilon.
\end{equation}
Since $Q$ is contractive, it follows that
\begin{equation}
\|F(T_i) - R \circ \mathbf{E} \circ Q(T_i) \|_\infty \leq \epsilon.
\end{equation}
Now apply the preceding result.
\end{proof}

Assume $H$ is a finite dimensional Hilbert space. We will consider
trace functionals on two distinct spaces of operators: one on the
space $\mathbf{L}(H)$, which we denote by $\opr{tr}$, and the other on
the space $\mathbf{L}(\mathbf{L}(H))$ of linear mappings
$\mathbf{L}(H) \rightarrow \mathbf{L}(H)$, which we denote
$\opr{tr}_\mathrm{oper}$.
We will prove that abelian factorizable completely positive maps cannot
approximate the identity map on $\mathbf{L}(H)$. To do this we will show that the trace
functional $\opr{tr}_\mathrm{oper}$ separates, in a sense to be made
precise in the next paragraph, the identity operator on
$\mathbf{L}(H)$ from abelian factorizable completely positive $F$.
Note that $\opr{tr}_\mathrm{oper}(I_{\mathbf{L}(H)}) =
\opr{dim}^2(H)$.

\begin{lem}\label{trace-separation-lemma} Suppose the Hilbert space
$H$ has finite dimension $n$. For any unit-preserving
abelian-factorizable completely positive map $F:\mathbf{L}(H)
\rightarrow \mathbf{L}(H)$,
\begin{equation}\label{separation-equation}
\opr{tr}_\mathrm{oper}(F) \leq n.
\end{equation}
\end{lem}
\begin{proof}
It suffices to prove this for $F$ which have the
form~\eqref{finite-factorization-equation}. Referring to that
representation, each positive functional $\rho_i$ can be represented
by a non-negative operator $S_i$ as follows:
\begin{equation}
\rho_i(T) = \opr{tr}(T S_i), 
\end{equation}
Since $\rho_i(I)=1$, $S_i$ also has unit trace and in particular,
$S_i\leq I$.  Moreover
$\sum_{i=1}^m G_i = I_H$. 
Now
\begin{equation}
\begin{aligned}
\opr{tr}_{\mathrm{oper}}(P)& = \sum_{i=1}^m \opr{tr}(S_i G_i) \\
& = \sum_{i=1}^m \opr{tr}(G_i^{1/2} S_i G_i^{1/2}) \\
& \leq \sum_{i=1}^m \opr{tr}(G_i^{1/2} G_i^{1/2}) \\
& = \sum_{i=1}^m \opr{tr}(G_i) = \opr{tr}(I_H) = n.
\end{aligned}
\end{equation}
\end{proof}
The previous lemma gives us a lower bound on the trace of $I -F$ for $F$ abelian factorizable:
\begin{equation}
\opr{tr}_\mathrm{oper}(I - F) \geq n^2 - n.
\end{equation}

We can use the above lower bound on the trace of $I -F$ to obtain a
lower bound on the norm of the operator $I - F:\mathbf{L}(H)
\rightarrow \mathbf{L}(H)$, where we consider $\mathbf{L}(H)$ with the
Schatten $2$-norm $\| \cdot\|_{2}$, defined in
Appendix~\ref{banach-spaces}.  The Schatten $2$-norm is the norm that
arises from the trace inner product on $\mathbf{L}(H)$, also known as
the Hilbert-Schmidt norm. We denote the corresponding operator norm on
$\mathbf{L}(H) \rightarrow \mathbf{L}(H)$ by $\| \cdot\|_{2\rightarrow
2}$.  

If $F$ is self-adjoint as an operator on the space $\mathbf{L}(H)$ with the
trace inner product,  then from Lemma~\ref{trace-separation-lemma}, we
immediately obtain the bound
\begin{equation} \label{2-2-norm-bounded}
\|I - F\|_{2\rightarrow 2} \geq 1 - \frac{1}{n}.
\end{equation} 
In fact, the lower bound~\eqref{2-2-norm-bounded} is true for general
unit-preserving abelian factorizable completely positive maps $F$. To
see this, write $F = F_\mathfrak{Re} + i F_\mathfrak{Im}$ where both
$F_\mathfrak{Re}$, $F_\mathfrak{Im}$ are self-adjoint operators (not
necessarily completely positive, however).  Now
\begin{equation}
\opr{tr}_\mathrm{oper}(I - F_\mathfrak{Re}) = \opr{tr}_\mathrm{oper}(I - F) \geq n^2 - n.
\end{equation}
Therefore 
\begin{equation}
\|I - F\|_{2\rightarrow 2} \geq \|I - F_\mathfrak{Re}\|_{2\rightarrow 2} \geq 1 - 1/n.
\end{equation}

In the discussion that follows, we need to consider another norm on
the space $\mathbf{L}(H) \rightarrow \mathbf{L}(H)$ in addition to the
norm $\| \cdot\|_{2\rightarrow 2}$ just considered.  The new norm,
which we denote $\| \cdot \|_{\infty \rightarrow \infty}$, is also an
operator norm on $\mathbf{L}(H) \rightarrow \mathbf{L}(H)$, but
relative to the $\| \cdot \|_\infty$ norm on $\mathbf{L}(H)$. The $\|
\cdot \|_{\infty \rightarrow \infty}$ norm is different from the $\|
\cdot\|_{2\rightarrow 2}$ norm, but for finite dimensional spaces $H$
the two norms are equivalent, which means that each norm is
bounded relative to the other.  To obtain the bounding constants, note
that if $T \in \mathbf{L}(H)$,
\begin{equation}
\|T\|_\infty \leq \|T\|_2 \leq \sqrt{n} \|T\|_\infty.
\end{equation}
From this it follows that 
\begin{equation}
\frac{1}{\sqrt{n}} \, \|F\|_{2\rightarrow 2}
\leq \|F\|_{\infty\rightarrow\infty} \leq \sqrt{n} \,
\|F\|_{2\rightarrow 2}.
\end{equation}
which is the desired relative bound.  In~\eqref{2-2-norm-bounded}, if
we substitute $F$ by $I - F$, we immediately obtain the following
proposition:
\begin{prop}\label{key-estimate}
Suppose $H$ is a Hilbert space of finite dimension $n$. If a
unit-preserving completely positive map $F:\mathbf{L}(H) \rightarrow
\mathbf{L}(H)$ has the form~\eqref{finite-factorization-equation}, then
\begin{equation}\label{minimum-conclusion}
\|I - F\|_{\infty\rightarrow\infty} \geq n^{-1/2}(1 - 1/n).
\end{equation}
\end{prop}
To prove the crucial result for the No-Go Theorem, we only use case
$n=2$ of~\eqref{minimum-conclusion}.
\begin{thm}\label{quantifiable-constraint-of-factorizable}
Suppose $H$ is of dimension $\geq 2$. If $F$ is a unit-preserving
completely positive map on $\mathbf{L}(H)$ with an abelian
factorization and $\beta < \sqrt{2}/4$, then
\begin{equation}\label{equation-of-approximation}
\|U^\dagger T U - F(T) \|_\infty \geq \beta \  \|T\|_\infty
\end{equation}
for at least one $T \in \mathbf{L}(H)$.  

If $H$ is finite dimensional, we can take $\beta = \sqrt{2}/4$.
\end{thm}
\begin{proof}
Replacing $F$ by the completely positive map 
$T \mapsto U F(T) U^\dagger$, we can assume without loss of generality that $U=I$.
To prove this, we will show that the assertion that 
\begin{equation} \label{wrong-assertion-to-be-refuted}
\|T - F(T) \|_\infty < \beta \ \|T\|_\infty, \quad \forall T \in \mathbf{L}(H),
\end{equation}
leads to a contradiction. However,~\eqref{wrong-assertion-to-be-refuted} implies
\begin{equation}\label{why-<-becomes-leq}
\| I - F \|_{\infty \rightarrow \infty} \leq \beta.
\end{equation}
We now reduce the proof to the case $H$ has dimension $2$, by
considering a Hilbert space $K$ of dimension $2$ and completely
positive unit-preserving mappings $Q:\mathbf{L}(K) \rightarrow
\mathbf{L}(H)$ and $R: \mathbf{L}(H) \rightarrow \mathbf{L}(K)$ such
that $R \circ Q$ is the identity map on $\mathbf{L}(K)$.
If $I_H$ can be approximated to within $\beta$ of the abelian factorizable
$F$, then $R  F  Q$ is also abelian factorizable and
\begin{equation}
\|I_K - R F Q\| = \|R I_H Q - R F Q\| \leq \|I_H - F\| \leq \beta < \sqrt{2}/4.
\end{equation}
However, in this contradicts Proposition~\ref{key-estimate}.

In the finite dimensional case, the norm is actually achieved so that
in~\eqref{why-<-becomes-leq} the $\leq$ sign can be replaced by $<$
and so we can take $\beta \leq \sqrt{2}/4$ as claimed.\footnote{
We note that it is possible to derive explicit results as well for the
case $n=3$. In this case one can show that one obtains a larger numerical bound than
$\sqrt{2}/4$, which applies when dim($H_{\mathrm{logical}})\geq 3$.}
\end{proof}

\subsection{Proof of Encoding No-Go Theorem}\label{ng_proof}

\subsubsection{Two Lemmas} \begin{lem}\label{triangle_lemma} Suppose
that $\mathbf{QCC}(P,U,\mathcal{M}_\mathrm{enc},
\mathcal{M}_\mathrm{dec},\alpha)$ holds.
If $P:\mathbf{T}(H_\mathrm{comp}) \rightarrow
\mathbf{T}(H_\mathrm{comp})$ is $\gamma$-damped and
$\mathfrak{A}$ and $\mathbf{E}_\mathfrak{A}$ are as identified
in Lemma~\ref{superop_gamma}, then for every $T
\in\mathbf{L}(H_\mathrm{logical})$,
\begin{equation} 
\|\mathcal{M}_\mathrm{enc}^\mathrm{t}   \mathbf{E}_\mathfrak{A} 
P^\mathrm{t} \mathcal{M}_\mathrm{dec}^\mathrm{t}(T)  - U^\dagger T  U \|_\infty
\leq (2 \gamma+\alpha) \|T\|_\infty. 
\end{equation}
\end{lem}
\begin{proof} The $\mathbf{QCC}(P,U,\mathcal{M}_\mathrm{enc},
\mathcal{M}_\mathrm{dec},\alpha)$ implies
that for every $\rho \in \mathbf{T}(H_\mathrm{logical})$ with
$\|\rho\|_1 \leq 1$ and self-adjoint $T \in
\mathbf{L}(H_\mathrm{logical})$,
\begin{equation}
\begin{aligned}
\biggl|\opr{tr}  \biggl[\biggl\{\mathcal{M}_\mathrm{dec} (P
(\mathcal{M}_\mathrm{enc}(\rho)))  - U \rho U^\dagger\biggr\} T\biggr]\biggr| 
& = \biggl|\opr{tr} \biggl[\rho \biggl\{
\mathcal{M}_\mathrm{enc}^\mathrm{t} (P^\mathrm{t}
(\mathcal{M}_\mathrm{dec}^\mathrm{t}(T))) - U^\dagger T U \biggr\} \biggr] \biggr| \\
& \leq \alpha \|T\|_\infty.
\end{aligned}
\end{equation}
It follows that for every $T \in \mathbf{L}(H_\mathrm{logical})$,
\begin{equation}
\|\mathcal{M}_\mathrm{enc}^\mathrm{t} (P^\mathrm{t}
(\mathcal{M}_\mathrm{dec}^\mathrm{t}(T))) - U^\dagger T U \|_\infty
\leq \alpha \|T\|_\infty.
\end{equation}
$\mathfrak{A}$ is commutative and by Lemma~\ref{superop_gamma}, for each $T \in
\mathbf{L}(H_\mathrm{logical})$ 
\begin{equation}
\|P^\mathrm{t} (T) - \mathbf{E}_\mathfrak{A} P^\mathrm{t}(T)\|_\infty \leq 2 \gamma
\|T\|_\infty.
\end{equation}
Thus using the fact that $\mathcal{M}_\mathrm{dec}$ and
$\mathcal{M}_\mathrm{enc}$ have norm $\leq 1$, for every $T \in
\mathbf{L}(H_\mathrm{logical})$, 
\begin{equation} \label{violated-condition}
\begin{aligned}
\|\mathcal{M}_\mathrm{enc}^\mathrm{t}  & \mathbf{E}_\mathfrak{A} 
P^\mathrm{t} \mathcal{M}_\mathrm{dec}^\mathrm{t}(T)  - U^\dagger T  U \|_\infty \\
& \leq \|\mathcal{M}_\mathrm{enc}^\mathrm{t} \mathbf{E}_\mathfrak{A}
 P^\mathrm{t} \mathcal{M}_\mathrm{dec}^\mathrm{t}(T) -
\mathcal{M}_\mathrm{enc}^\mathrm{t} P^\mathrm{t}\mathcal{M}_\mathrm{dec}^\mathrm{t}(T)\|_\infty \\
& + \|
\mathcal{M}_\mathrm{enc}^\mathrm{t}
P^\mathrm{t}\mathcal{M}_\mathrm{dec}^\mathrm{t}(T) - U^\dagger T U
\|_\infty \\
& \leq (2 \gamma+\alpha) \|T\|_\infty. \end{aligned}
\end{equation}
\end{proof}
\begin{lem}\label{fac_lem}
If $H_{\mathrm{logical}}$ is of dimension $\geq 2$, and $P$, $U$, $\mathcal{M}_\mathrm{enc}$,
$\mathcal{M}_\mathrm{dec}$ and $\mathbf{E}_\mathfrak{A}$
are the same as in Lemma \ref{triangle_lemma}, and if $\beta <  \sqrt{2}/4$, 
then for some non-zero $T \in \mathbf{L}(H_\mathrm{logical})$,
\begin{equation}\label{special_number}
\| \mathcal{M}_\mathrm{enc}^\mathrm{t} \mathbf{E}_\mathfrak{A} 
P^\mathrm{t} \mathcal{M}_\mathrm{dec}^\mathrm{t}(T)  - U^\dagger T  U \|_\infty
\geq \beta\| T \|_\infty
\end{equation}
\end{lem}
\begin{proof}
The unit-preserving completely positive map
$R=\mathcal{M}_\mathrm{enc}^\mathrm{t} \mathbf{E}_\mathfrak{A}
P^\mathrm{t} \mathcal{M}_\mathrm{dec}^\mathrm{t}$ factors through the
abelian von Neumann algebra $\mathfrak{A}$; this follows from the
presence of the projection operator $\mathbf{E}_\mathfrak{A}$ in the
expression for $R$. Then \eqref{special_number} follows from
Theorem \ref{quantifiable-constraint-of-factorizable}.
\end{proof}

\subsubsection{Statement of Proof of Encoding No-Go Theorem}
\begin{proof}
By the hypotheses of the Encoding No-Go Theorem (Theorem \ref{ngt1}),
the quantity $2 \gamma+\alpha < \sqrt{2}/4$. Choose $\beta$ such that
$2 \gamma + \alpha < \beta < \sqrt{2}/4$.  From Lemma
\ref{triangle_lemma}, it follows that for every $T \in
\mathbf{L}(H_\mathrm{logical})$,
\begin{equation}\label{first-step-in-proof-by-contradiction-of-no-go}
\| \mathcal{M}_\mathrm{enc}^\mathrm{t} \mathbf{E}_\mathfrak{A}
P^\mathrm{t} \mathcal{M}_\mathrm{dec}^\mathrm{t}(T) - U^\dagger T U
\|_\infty < (2 \gamma + \alpha) \| T \|_\infty.  
\end{equation}
On the other hand by \eqref{special_number} in Lemma \ref{fac_lem},
there is a non-zero $T$ such that
\begin{equation}\label{second-step-in-proof-by-contradiction-of-no-go}
\| \mathcal{M}_\mathrm{enc}^\mathrm{t} \mathbf{E}_\mathfrak{A}
P^\mathrm{t} \mathcal{M}_\mathrm{dec}^\mathrm{t}(T) - U^\dagger T U
\|_\infty \geq \beta \| T \|_\infty.  
\end{equation}
Since $\| T\|_\infty >  0$,
equations~\eqref{first-step-in-proof-by-contradiction-of-no-go}
and \eqref{second-step-in-proof-by-contradiction-of-no-go}
imply $\beta \leq 2 \gamma + \alpha$, which contradicts the
choice of $\beta$.
\end{proof}

\subsection{Interpretation of Encoding No-Go Theorem}

The Encoding No-Go Theorem is an extremely powerful application of the QCC.
With this theorem, one can calculate the amount of damping for which
fault-tolerant quantum computation becomes impossible.
When the amount of damping exceeds the critical amount, which means that the
value of $2\gamma$ becomes less than the critical
value $2\gamma_{\mathrm{critical}} = \sqrt{2}/4 - \alpha$, we find
that the only solutions to the QCC are those for which dim $H_{\mathrm{logical}}=1$.
As noted above, in this case no meaningful quantum computation is possible, since
not even one quantum bit can be accomodated.

In order to analyze the constraints on solutions to
$\mathbf{QCC}(P,U,\mathcal{M}_\mathrm{enc},\mathcal{M}_\mathrm{dec},\alpha)$
implied by the No-Go Theorem,
we regard the pair $\{ U, \alpha\}$ as given, since both the
desired quantum computation $U$, and the maximum acceptable implementation
inaccuracy $\alpha$, are prescribed.
We assume that $U$ is defined on a Hilbert space $H_{\mathrm{logical}}$ such that
${\mathrm{dim}}(H_{\mathrm{{logical}}})\geq 2$, to allow meaningful quantum
computation.
For practical applications, one then seeks to determine triples
$\{ P,\mathcal{M}_\mathrm{enc},\mathcal{M}_\mathrm{dec}\}$
that satisfy $\mathbf{QCC}(P,U,\mathcal{M}_\mathrm{enc},\mathcal{M}_\mathrm{dec},\alpha)$.
The No-Go Theorem, on the other hand,
identifies conditions in which
$\mathbf{QCC}(P,U,\mathcal{M}_\mathrm{enc},\mathcal{M}_\mathrm{dec},\alpha)$
is not satisfied.
Thus the No-Go Theorem provides a useful calculational tool for eliminating prospective
quantum computer implementations that are guaranteed to fail.
With the criterion provided by the No-Go Theorem, we can bound the space of
solutions to $\mathbf{QCC}(P,U,\mathcal{M}_\mathrm{enc},\mathcal{M}_\mathrm{dec},\alpha)$.
This allows exploration of trade-offs amongst the members of the triple
$\{ P,\mathcal{M}_\mathrm{enc},\mathcal{M}_\mathrm{dec}\}$, with which one
may construct generalized ``phase diagrams" that indicate boundaries between
{\em potentially acceptable}
and {\em definitely unacceptable} values for $P$, $\mathcal{M}_\mathrm{enc}$ and
$\mathcal{M}_\mathrm{dec}$.

\section{Quantum Components}\label{qm_components}

\subsection{Introduction}

As defined in Section \ref{QCC_section}, we refer to a physically realizable device
intended to implement a quantum computation as a {\em quantum component.} Mathematically,
a quantum component is represented by a completely positive, trace preserving map, $P$.
The detailed form of $P$, as an explicit function, is dictated by underlying
equations of motion. 

For a closed physical system, the quantum mechanical dynamics of the system
are given by the Schr\"odinger
equation associated to a particular Hamiltonian
$\mathcal H$. However, for the general problem of a {\em practical}
quantum computer, we must analyze realistic quantum
components that interact with their environment, dissipate heat and exhibit decoherence.
We must thus utilize a formulation that yields equations of motion
appropriate to open quantum mechanical systems.

The connection
to the Quantum Computer Condition presented in Section \ref{QCC_section} is made by
starting with the appropriate equations of motion that describe the dynamical
evolution of a realistic quantum component interacting with its environment.
One proceeds by solving the appropriate equations of motion. The explicit
solution thus obtained
provides the time-dependence of the quantum mechanical state of the open
system. In principle, this
allows us to deduce the explicit functional form of $P$.

Formulations of equations of motion for quantum components
that interact with complex
environments comprised of many degrees-of-freedom
necessarily involve approximations of one sort or another, and there is not
in general a unique choice. In this section we illustrate the general approach by
making use of a Lindblad-type equation (made more precise below) to describe how
one obtains the quantum component $P$ that appears in the Quantum Computer Condition.

\subsection{Dynamical Equations of Motion}

\subsubsection{Time-dependent generalization of the Lindblad equation} The state of a
quantum mechanical system defined on a Hilbert space $H$
can be modeled by a density operator $\rho(t)$ whose time dependence obeys a
linear (time dependent) equation
\begin{equation} \label{generic-equation-of-evolution}
\frac{d}{dt} \rho (t) = A(t) \rho(t),
\end{equation}
where $A(t)$ is an operator acting on $\mathbf{T}(H)$, the Banach space of trace class
operators on $H$.
For a closed system we have the Schr\"odinger equation, and
the right-hand-side of \eqref{generic-equation-of-evolution} is given by 
\begin{equation} \label{SchrodingerOperator}
A(t) \rho = -\frac{i}{\hbar} [\mathcal{H}(t), \rho],
\end{equation}
where the Hamiltonian $\mathcal{H}(t)$ is a self-adjoint operator which may depend
on the parameter $t$.

However, it would be impractical to describe realistic quantum components using
\eqref{SchrodingerOperator}, since writing down the detailed Hamiltonian operator
to account for all of the degrees-of-freedom comprising the quantum component
and its environment would be intractable.

Motivated by the work of Lindblad~\cite{lindblad76}, we make use instead of the following
expression for the action of $A(t)$ on $\rho$:
\begin{equation} \label{LindbladOperator}
A(t) \rho = -\frac{i}{\hbar} [\mathcal{H}(t), \rho] + \sum_j \bigg[ L_j(t)
\rho L_j^\dag(t) - \frac{1}{2}\big\{L_j^\dag(t) L_j(t), \rho\big\}\bigg]
\end{equation}
where as above the $\mathcal{H}(t)$ is a self-adjoint operator which may depend
on the parameter $t$, and the $L_j$ are operators
that describe effects arising from interaction with the environment, such as dissipation
and decoherence.
These operators are generalizations of the Lindblad operators
of~\cite{lindblad76}; unlike the treatment given by Lindblad in~\cite{lindblad76},
however,
which considered only time-independent Hamiltonians $\mathcal{H}$ and
time-independent dissipative perturbations $L_j$, with all operators bounded,
we will allow unbounded and
time-dependent $\mathcal{H}(t)$ and time-dependent (but still bounded) dissipative
perturbations $L_j(t)$. Our treatment generalizes that given in~\cite{lindblad76},
and we will refer to the equation of evolution
\begin{equation} \label{LindbladEquation}
\frac{d}{dt}\rho(t) =  -\frac{i}{\hbar} [\mathcal{H}(t), \rho] + \sum_j \bigg[ L_j(t)
\rho L_j^\dag(t) - \frac{1}{2}\big\{L_j^\dag(t) L_j(t), \rho\big\}\bigg]
\end{equation}
as the {\em time-dependent generalization of the Lindblad equation}.
As an example of how one may approximate
the dynamics of a quantum component interacting with a complex environment,
this equation provides a starting point to the derivation of $P$ used in the
Quantum Computer Condition.
For this purpose we need to consider solutions to equations of the type given in
\eqref{LindbladEquation} known as ``fundamental solutions."

\subsection{Fundamental Solutions}

The concept of a {\em fundamental solution} associated to an evolution equation
(see~\cite{tanabe}, \S4.4) of the
general form~\eqref{generic-equation-of-evolution}, such
as \eqref{LindbladEquation} in particular,
where $\rho$ is a function with values in the Banach space
$\mathbf{T}(H)$ and $A(t)$ is a one-parameter family of (possibly
unbounded) linear operators on $\mathbf{T}(H)$, will play an important
role in this paper.
A {\em fundamental solution} associated to an equation of motion is a solution of an
operator version of the original equation. We show below how, given a fundamental
solution, one obtains the completely positive, trace preserving map $P$ that appears
in the Quantum Computer Condition.

Fundamental solutions are given by a family
$\{P_{t,s}\}_{t \geq s \geq 0}$
of bounded operators on $\mathbf{T}(H)$ indexed on pairs of real
numbers $t$ and $s$ satisfying equations
\begin{equation}\label{fundamental-solution-def}
\frac{d}{dt}P_{t,s} = A(t) P_{t,s} 
\end{equation}
and
\begin{equation}
P_{s,s} = I.
\end{equation}
The intended
interpretation of $P_{t,s}$ is that if the system is in state
$\rho$ at time $s$, then the system will be in state $P_{t,s}\cdot\rho$
at later time $t$, that is
\begin{equation}\label{evolve}
\rho(t) = P_{t,s}\cdot\rho(s).
\end{equation}

\subsubsection{Existence and Positivity Properties of Fundamental Solutions}

The problem of the existence and uniqueness of fundamental solutions is a
central one in the mathematical theory of evolution equations.
In addition to addressing the question of the existence of fundamental solutions, it is
important for our analysis to determine the positivity properties of
fundamental solutions.
This is because, as we shall see,
the complete positivity of the map $P$ that explicitly
appears in the Quantum Computer Condition is inherited from
the complete positivity of the set of fundamental solutions $\{P_{t,s}\}$.
Positivity is important in ensuring
that the set $\{P_{t,s}\}$, as well as $P$, carry density matrices to density
matrices.

For equations of motion associated to
finite dimensional systems, existence and uniqueness of solutions
follows from the Lipschitz theorem on ordinary differential
equations.
Lindblad's analysis extended to infinite-dimensional (but bounded) systems, so,
more generally, Lindblad characterized the infinitesimal
generator of a {\em norm continuous} completely positive
semigroup~\cite{lindblad76},
which corresponds to the case of~\eqref{generic-equation-of-evolution}
in which $A(t)$ is constant and norm bounded
(but possibly infinite-dimensional), {\em i.e.}, for the standard,
time-independent Lindblad equation.

In order to generalize Lindblad's analysis to allow us to study
infinite-dimensional, unbounded,
time-dependent quantum systems, we will need to consider general evolution
equations~\eqref{generic-equation-of-evolution} in which $A(t)$ may be
unbounded as well as time-dependent, for the infinite-dimensional case.
The general theory of such
equations was developed by Kato, Yosida and others in the 1950's.
We will rely on results of Kato~\cite{kato} which pertain to both existence of solutions
and positivity properties, and on Theorem
4.4.1 of~\cite{tanabe} which pertains to existence of solutions;
moreover, we will use a constructive form of
the theorem (which follows from an examination of the proof) which
expresses the fundamental solution as a limit of a product of
exponentials. The basic assumption of the approach is that if the $A(t)$ are generators
with sufficiently smooth variation, then fundamental solutions exist.

Our system comprised of a quantum component interacting with its environment,
described by \eqref{LindbladEquation}, consists of a time-dependent
Hamiltonian $\mathcal{H}(t)$ characterized by a time-dependent perturbation 
$V(t)$ of a (possibly unbounded) self-adjoint operator $\mathcal{H}_0$ so that
we have $\mathcal{H}(t) = \mathcal{H}_0 + V(t)$. In order to
apply Kato's theory for time dependent evolution equations, we will
assume among other things that the perturbation $V(t)$ does not change the
domain of $\mathcal{H}_0$.  For the necessary background
see~(\cite{kato},\cite{tanabe}).
We now state a proposition that asserts the existence and complete positivity
of fundamental solutions
to operator versions of the time-dependent generalization of
the Lindblad equation given in \eqref{LindbladEquation}:

\begin{prop}
For suitably regular time-varying potentials $V(t)$ and dissipation
operators $L_j(t)$, there exists a strongly continuous completely positive
operator $P_{t,s}$ which is a fundamental solution to the time-dependent
generalization of the Lindblad equation.
\end{prop}
The exact statement of the conditions for the above in the form of
a theorem, and proof, are given in Appendix \ref{mathematical-preliminaries-section}
in \S\ref{solve_lindblad} and \S\ref{existence-and-uniqueness}.

\subsection{Construction of Quantum Components}

Having established the existence and complete positivity of
fundamental solutions to the operator
form of the underlying equation of motion for our system (comprised of
the quantum component interacting with its environment),
it is straightforward to
obtain an expression for the completely positive, trace preserving map $P$,
characterizing the quantum component, that explicitly
appears in the QCC. This is simply obtained by noting ({\em cf} \eqref{evolve}) that the
time-evolution of the state $\rho(t)$ between an initial fixed
time $\hat s$ (corresponding to
the start of the quantum computation) and a final fixed time $\hat t$ (corresponding to the
end of the quantum computation) is fully specified by
the fundamental solution defined with respect to those time values,
$P_{\hat {t}, \hat s}$, so that we have
\begin{equation}
\rho(\hat t ) = P_{\hat {t}, \hat s}\cdot \rho(\hat s ),
\end{equation}
and hence the completely positive, trace-preserving map $P$ that appears
in the QCC is given by the equivalence
\begin{equation}\label{qcomponent_def}
P\equiv P_{\hat {t}, \hat s}~.
\end{equation}

\section{Unified Treatment of Quantum Computing Paradigms}\label{unification}

\subsection{Introduction}
In this section we show that the QCC provides a unifying framework in which to
describe on the same footing the currently-known ``paradigms" for quantum computation,
including the {\em circuit-based paradigm}, the {\em graph state-based paradigm},
the {\em adiabatic quantum computer paradigm}.
The QCC subsumes all of these into a single, unifying paradigm for quantum computing.

\subsection{Circuit-based paradigm}

In this section we describe the specification of quantum components based on
the ``circuit-based" 
paradigm of quantum computation.
We proceed as follows:\\
{\bf{(1)}} For purposes of clarity, we
begin in \ref{ideal_circuit}
by working in an idealization in which there is no noise present.
For this idealized case we obtain the general form
of the completely positive 
map $P$ characterizing the quantum component.
We then apply the result (for the noiseless idealization)
to several specific realizations of the
circuit-based paradigm.
These include
qubit-based quantum computers (these utilize states, the operators for which have
a discrete eigenspectrum, {\em i.e.}, qubits in the case of 2-level systems),
quantum continuous variable-based quantum computers (these utilize states, the
operators for which have
a continuous eigenspectrum, such as coherent states),
and liquid state NMR-based quantum computers.\\
{\bf{(2)}} Having obtained the general form for $P$ in the noiseless case, for each of the
three above-mentioned realizations of the circuit-based paradigm, we then explain
in \ref{noise_circuit} how
to modify the analysis, in a way appropriate to all choices of circuit
realization, so as to account for the effects of noise.

\subsubsection{Idealized Circuits in the Absence of Decoherence and Dissipation}
\label{ideal_circuit}
A quantum circuit is described by a set $G$ of gates operating 
in some specified order on elements of 
a set $R$ of objects.  The quantum states of these objects constitute the information which 
is ``processed" 
by the gates of the quantum circuit.  
We associate 
with each object $i$ of $R$ a Hilbert space $H^{(i)}$ that describes the possible states 
of that particular object.  The Hilbert space for the full set of objects on which the circuit 
operates is then

\begin{equation}
H_{\mathrm{circuit}} = \bigotimes_{i \in R} H^{(i)}~.  
\end{equation}

\noindent The gates in $G$, labeled by the index $\mu$,
are described by unitary operators $\hat{V}_\mu$, so that 
the unitary operator describing the (noiseless) operation of the circuit is  

\begin{equation}
\label{EQ:vcircuit}
{V}_{\mathrm{circuit}} = \prod_{\mu \in G} \hat{V}_{\mu}~,   
\end{equation}

\noindent where the product of operations is ordered in accordance
with the definition of the circuit.
The factors $\hat{V}_{\mu}$ that appear in the multiplicand of \eqref{EQ:vcircuit}
are in principle obtained from the fundamental solution to the
appropriate underlying equation of motion, following
the procedure outlined in Section \ref{qm_components} above.\footnote{It is
extremely important to note that we
are {\em not} discussing here the abstractly defined quantum
computation itself, which is prescribed in advance, and is
represented by the unitary operator $U$ that appears explicitly
in the second term under the norm symbol in the QCC given in \eqref{encoded_QCC}.
Rather, we are discussing quantities
that represent elements of a physical device that is to be used as an actual
quantum computing machine. As such, the quantities under discussion here are to be regarded
as ``building blocks" for the {\em first} term under the norm symbol in
\eqref{encoded_QCC}\label{fnote_P_not_U}.}

Each gate operates on a subset $\Sigma_\mu$ of the information elements, leaving 
the rest unaffected:

\begin{equation}
\label{EQ:vsigmamu}
\hat{V}_{\mu} = V_\mu^{\Sigma_\mu} \otimes
{\Big(}\bigotimes_{i \notin \Sigma_\mu} I_{H^{(i)}}{\Big)}~,
\end{equation}

\noindent where $I_{H^{(i)}}$ is the identity operator on $H^{(i)}$,
and $V_\mu^{\Sigma_\mu}$ is the transformation effected by the $\mu$th gate,
so that we may express the unitary operator describing the idealized circuit as 

\begin{eqnarray}
\label{EQ:pcircuit}
P\cdot\rho &=& V_{\mathrm{circuit}} \rho V_{\mathrm{circuit}}^\dag \\
      &=& \left( \prod_{\mu \in G} \hat{V}_{\mu} \right) \rho 
          \left( \prod_{\mu \in G}^\prime \hat{V}_{\mu}^\dag \right)~, 
\end{eqnarray}

\noindent where the prime indicates that the second product reverses the order of the 
factors relative to the first product.\footnote{We
strongly reiterate the message given in
Footnote \ref{fnote_P_not_U}: It is only coincidentally the case that the form of
\eqref{EQ:pcircuit} resembles that of \eqref{ersatz-quantum-computer-equation}.
{\em Both} the right- and left-hand sides of \eqref{EQ:pcircuit} describe the
physical device ({\em {i.e.}}, $P$), not the abstractly-defined quantum computation $U$,
and thus both sides of eq.(\ref{EQ:pcircuit}) are to be
associated with the first term under the norm symbol in the QCC given
in \eqref{encoded_QCC}.\label{reminder}}


Up to this point we have made no restrictions on the Hilbert spaces or 
the types of gates appearing in the specification of the circuit. We now
describe three special cases of interest within the circuit-based
paradigm: qubits, quantum continuous 
variables, and liquid state NMR.
We obtain the completely positive, trace preserving map $P$ that characterizes the 
implementation of the circuit for each example, thus showing that the QCC provides
the proper foundational presentation of a quantum computer for all the cases.

{\noindent{ { {(Case 1)}} {\em {Circuit-realization with qubits}}}}

The great majority 
of the research in quantum computation has focussed on circuits for which the information 
elements are qubits defined on a two dimensional Hilbert space $\mathbb{C}^2$, so that the 
full Hilbert 
space of the circuit is 

\begin{equation}
H_{\mathrm{circuit}} = \left (\mathbb{C}^2 \right)^{\otimes |R|}~.  
\end{equation}
where $|R|$ is the cardinality of the set of computational qubits.

Usually the set of gates is chosen from a relatively small set of operations, each 
of which acts on only 1 or 2 qubits. Specializing \eqref{EQ:vsigmamu} to
the case of a circuit
built out of gates acting only
on 1 or 2 qubits (this would be the proper description, for
instance, of a machine that uses the universal set of quantum gates), we have

\begin{equation}
\hat{V}_{\mu} = V_{\mu}^{(i)}\otimes \bigotimes_{k (\neq i) {\in R}} I_k,
\end{equation}
or
\begin{equation}
\hat{V}_{\mu} = V_{\mu}^{(ij)}\otimes \bigotimes_{k (\neq i,j)
{\in R}} I_k~.  
\end{equation}
where $I_k$ is the identity operator acting on the copy of $\mathbb{C}^2$ associated
to the $k$th qubit.
With these definitions, the operator $P$ that characterizes the implementation of
the qubit-based realization of a
quantum computer designed according to the circuit-based paradigm is
given by (\ref{EQ:vcircuit}) and (\ref{EQ:pcircuit}). We thus see that the
circuit-based paradigm, on which a large amount of the research in the field
is based, is properly described by the QCC.

\noindent{{ {(Case 2)}} {\em {Circuit-realization with quantum continuous variables}}}

Alternatively, we may select different Hilbert spaces $H^{(i)}$ appropriate for  
quantum computation using quantum continuous variables (QCV).  For instance, 
the $H^{(i)}$ might describe the  
states of simple harmonic 
oscillators.  The full Hilbert space of the circuit, and the gate operations out of which it
is built, are then defined 
analogous to the above prescriptions for the qubit-based quantum computer.   
In this way we arrive at a specification
of the quantum component $P$ appropriate to the case of computation by QCV.
Thus, the QCV realization of the circuit-based paradigm is also seen to be
properly described by the QCC.

\noindent{{ {(Case 3)}} {\em {Circuit-realization with NMR states}}}

The treatment of quantum computation by nuclear magnetic resonance (NMR) in liquids 
requires special consideration due to the fact that the NMR sample effectively contains 
many copies of the circuit carrying out the same computation.  In this case we define the 
Hilbert space of the system as

\begin{equation}
H_{\mathrm{NMR}} = H_{\mathrm{circuit}}^{\otimes N_s}~,
\end{equation}

\noindent where $N_s$ is the number of copies of the circuit,
which requires that we extend the definition of the unitary operation for 
the circuit as follows:

\begin{equation}
\label{EQ:vnmr}
{V}_{\mathrm{NMR}} = {V}_{\mathrm{circuit}}^{\otimes N_s} ~.
\end{equation}

\noindent We then obtain 

\begin{equation}\label{eq_nmr}
P\rho = {V}_{\mathrm{NMR}} \rho {V}_{\mathrm{NMR}}^\dag
\end{equation}

\noindent which describes the NMR-based quantum computer in the QCC.\footnote{In
connection with \eqref{eq_nmr} we
repeat the admonition given in Footnote \ref{reminder} .}

\subsubsection{Quantum Circuits in the Presence of Decoherence and Dissipation}
\label{noise_circuit}

Thus far in this section we have restricted ourselves to a discussion of quantum circuits 
in the absence of decoherence and dissipation.  
This is reflected in our description of the 
gates as implementing purely unitary operations, as in (\ref{EQ:vcircuit}).  Even at this level of 
description the circuit is not necessarily error-free.  Unitary errors derive from a situation in 
which the evolution of the quantum circuit is unitary, but the circuit does not implement exactly 
the desired unitary computation:

\begin{equation}
\label{EQ:imperfectcircuit}
V_{\mathrm{circuit}} \neq U~.  
\end{equation}

\noindent Unitary errors can arise from either the design or the physical 
implementation of the circuit.  
Design errors arise, for example, due to the the fact that a universal set of 
quantum gates only allows 
for the implementation of an arbitrary unitary operation U to within an arbitrarily small 
tolerance~\cite{nielsen-chuang}.  
In general there will then be some residual error implied by the very design of the circuit.  
Implementation errors result from inaccuracies in the 
physical parameters governing the unitary evolution associated with a gate as compared with 
the specification of those parameters by the design.  
For example, an interaction 
Hamiltonian may be applied for a longer time than specified, or there may be errors in the 
field strengths or couplings in the interaction Hamiltonian.  

In addition to unitary errors, we also need to deal with errors resulting from decoherence and 
dissipation.   
We will refer to these simply as decoherent errors.\footnote{The case of liquid state 
NMR admits another type of error, which arises when 
the operators $\hat{V}_{\mathrm{circuit}}^{\left( k \right)}$ appearing in (\ref{EQ:vnmr}) 
effect different unitary errors 
on different copies ($k$) of the circuit.  These are known as incoherent errors.}
In this case the gates are described 
by completely positive maps $\hat{P}_{\mu}$,
where the hat on the $P$ indicates that this quantum component is a single gate,
and the index $\mu$ identifying the particular gate, as above.
In addition to replacing unitary operations representing the gates 
by completely positive maps, it is also 
important to account for errors occurring in the transmission of quantum states from 
one gate to the next.    
This means that the transmission channels are also represented by 
completely positive maps designated by the symbol $\hat{P}_{\mu}$.  
In effect, the 
transmission channels (including any quantum memories used to store the states) 
are regarded as gates that (ideally) implement the identity transformation:
$\hat{V}_\mu = I$.  If we call the set of transmission channels $C$, the completely 
positive map that describes the circuit is then

\begin{equation}
\label{EQ:prodpmu}
P = \prod_{\mu \in G \cup C} \hat{P}_{\mu}~,  
\end{equation}

\noindent where the index $\mu$ 
now identifies both the transmission channels (in $C$) and the gates (in $G$). 

The operators $\hat{P}_\mu$ are obtained by explicitly solving the 
equations of motion for the physical device that implements the gate.  The result may 
be written formally as the sum of two terms, one of which represents the action of the 
unitary operator $\hat{V}_\mu$ describing the ideal gate as specified by the circuit design, 
and the other of which represents the effects of 
unitary implementation errors as well as
decoherence and/or dissipation:

\begin{equation}
\label{EQ:gateerrormodel}
\hat{P}_\mu \rho = \left( 1 - \epsilon_\mu \right) 
   \left( \hat{V}_\mu \rho \hat{V}_\mu^\dag \right) + 
   \epsilon_\mu \hat{Q}_\mu \rho ~,
\end{equation}

\noindent where $\epsilon_\mu$ is the probability that an error occurs during the
operation of the 
gate, and $\hat{Q}_\mu$ is a completely positive map that represents the effects of the 
error. Although we have indicated how this follows in principle
from an explicit solution of the detailed 
dynamics of the gate (as described in \S\ref{qm_components}),
an error model of this form is often assumed from the outset.  The 
latter approach necessitates the choice of some specific 
operator $\hat{Q}_\mu$ to represent the errors.  For instance,
in investigating the properties of quantum error correcting codes one often invokes a
``depolarization" qubit error model in which

\begin{equation}
\label{EQ:depol}
\hat{Q}_\mu \rho \equiv {1\over 4}\left( \rho + \sum_{j=1}^3 \sigma_j \rho
\sigma_j\right)~,
\end{equation}

\noindent where the  $\sigma_j$ are the Pauli matrices acting on a single qubit.
The advantage of this approach is that 
it abstracts the physical implementation of the quantum computer from the design of the 
circuit while retaining the main features of decoherence that must be addressed in the 
development of any practical quantum computer.  The disadvantage is that the abstraction 
must then be justified relative to the detailed physical implementation of 
the computer in order 
to apply rigrously the results of any analysis based on \eqref{EQ:depol}.  

Whether we arrive at (\ref{EQ:gateerrormodel}) by detailed calculation or by abstraction, 
we may use it in conjunction with (\ref{EQ:prodpmu}) to describe the operation of the 
entire circuit:

\begin{equation}
P\cdot\rho = \left[ \prod_\mu \left( 1 - \epsilon_\mu \right) \right]
         V_{{\mathrm{circuit}}} \rho V_{{\mathrm{circuit}}}^\dag +
  \left\{ 1 - \left[ \prod_\mu \left( 1 - \epsilon_\mu \right) \right] \right\}
         \epsilon_f Q_f \rho~,
\end{equation}

\noindent where  
$Q_f$ incorporates the effects of the individual gate errors.  We define
an error probability associated with the entire circuit:

\begin{equation}\label{eps_mu_f}
\epsilon_f \equiv 1 - \prod_\mu \left( 1 - \epsilon_\mu \right)~,
\end{equation}

\noindent with which we have 

\begin{equation}
\label{EQ:circuiterrormodel}
P\cdot\rho = \left( 1 - \epsilon_f \right)
         V_{{\mathrm{circuit}}} \rho V_{{\mathrm{circuit}}}^\dag +
  \epsilon_f Q_f \rho ~.  
\end{equation}
We will make use of this error model in applying the QCC to the problem of 
finding error thresholds for fault tolerance in \S\ref{et_ft}.

We note that eqs. \eqref{EQ:gateerrormodel} and \eqref{EQ:circuiterrormodel}
connect theoretical analysis with experimental observations.
A general theoretical method for obtaining an explicit
expression for the quantum component, $P$, in terms of the underlying equation
of motion for the system, has been given above in \S\ref{qm_components}. From an
{\em experimental} perspective, explicitly writing
\eqref{EQ:gateerrormodel} or \eqref{EQ:circuiterrormodel}
is a goal of quantum process tomography,
which is by definition
the experimental method of determining the evolution of open
quantum systems ~\cite{YSW}.

\subsection{Adiabatic Quantum Computing Paradigm}

We now show that the QCC also provides the proper framework in which to formulate
the adiabatic quantum computing paradigm.
In adiabatic quantum computing, the quantum component can be described by a Lindblad 
equation incorporating a Hamiltonian of the form:

\begin{equation}
\mathcal{H}_{\mathrm{adiabatic}} \left( t \right) = f \left( t\right) \mathcal{H}_0 + 
      g \left( t\right) \mathcal{H}_f
\end{equation}

\noindent where $f$ and $g$ are smooth functions of time with 
$f(0) = 1$, $f(T) = 0$, $g(0) = 0$ and $g(T) = 1$, so that the adiabatic Hamiltonian 
goes smoothly from $\mathcal{H}_0$ to $\mathcal{H}_f$ over time.  
The full Lindblad equation may then be written as
\begin{equation} \label{EQ:adiabaticlindblad}
\frac{d}{dt} \rho (t) = -\frac{i}{\hbar} 
[\mathcal{H}_{\mathrm{adiabatic}}(t) + \mathcal{V}(t), \rho] + 
\sum_j \bigg[ L_j(t) \rho L_j^\ast(t) - 
\frac{1}{2}\big\{L_j^\ast(t) L_j(t), \rho\big\}\bigg]~,
\end{equation}
where $\mathcal{V}(t)$ represents unitary errors and the terms involving $L_j(t)$ 
represent interactions with the environment.  

The computation begins with an an initial preparation of the ground 
state of the Hamiltonian $\mathcal{H}_0$.  
Provided the conditions for adiabatic 
evolution are satisfied, evolution under the exact adiabatic Hamiltonian takes the 
ground state of $\mathcal{H}_0$
to the ground state of $\mathcal{H}_f$ 
with a high degree of accuracy.  
We identify the operator
$U$ that appears in the QCC as a unitary operator that describes this desired behavior:

\begin{equation}
U: |\phi_0 \rangle \mapsto |\psi_0 \rangle ~.  
\end{equation}

\noindent where $|\phi_0 \rangle$ is the ground state of $\mathcal{H}_0$, 
$|\psi_0 \rangle$ is the ground state of $\mathcal{H}_f$, and $U$ is otherwise arbitrary.  

The physical realization 
of the adiabatic quantum computer is described by 
(\ref{EQ:adiabaticlindblad}), and thus implements the desired operation $U$ only approximately.  
This is due to the approximation inherent in the adiabatic evolution
itself, as well as any additional 
unitary errors, decoherence and dissipation.  
We express this 
fact quantitatively by using the fundamental solution of 
(\ref{EQ:adiabaticlindblad}) to derive the map $P$
(as described above in \S\ref{qm_components})
that describes the operation of the 
adiabatic quantum component. The construction of the QCC then follows.  

Note that the special case of the adiabatic 
quantum computer is characterized under $\mathbf{QCC}(P,U,\mathcal{M}_\mathrm{enc},
\mathcal{M}_\mathrm{dec},\alpha)$
by two unique features: 

\begin{enumerate}
\item the encoding and decoding maps, $\mathcal{M}_\mathrm{enc}$ and 
$\mathcal{M}_\mathrm{dec}$ are identities, and
\item the QCC is required to hold only for the ground state of the initial 
Hamiltonian, that is, for $\rho = |\phi_0 \rangle \langle \phi_0 |$ ~.  
\end{enumerate}

\subsection{Graph state-based (including cluster state-based) paradigm}

\begin{figure}
\input{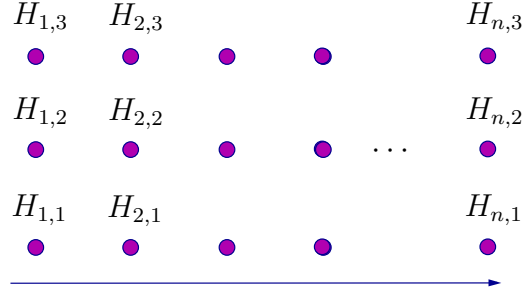}
\caption{Two-dimensional cluster configuration (the arrow denotes time flow)}
\label{cluster-configuration-figure}
\end{figure}

We now consider the cluster-based approach to quantum
computation~(\cite{raussendorf-briegel-2002},\cite{raussendorf-briegel-2001},
\cite{raussendorf-browne-briegel-graph-states}),
which is formulated in terms of an array of two level quantum systems.  The
systems in the array are referred to as sites and are elements of some
set $L$ which has a geometrical structure such as a 1 or 2 dimensional
lattice; in addition to the geometrical structure there is also a flow
structure which models the flow of information through the
cluster. The two-dimensional Hilbert space corresponding to a site
$\mathbf{s} \in L$ is denoted $H_\mathbf{s}$ and the Hilbert space $H$
of the entire cluster system is the tensor product
\begin{equation}
H_\mathbf{cluster} = \bigotimes_{\mathbf{s} \in L} H_\mathbf{s}.
\end{equation}
A key concept in the cluster approach is {\em measurement at a site}
$\mathbf{s} \in L$ considered as an operation on quantum mechanical
states.  As an operation on pure states, the measurement corresponds
to a pair of self-adjoint projections $E_\mathbf{s}$, $1-
E_\mathbf{s}$ on $H_\mathbf{s}$.  In terms of the cluster system, we
identify the projection $E_\mathbf{s}$ with a projection on
$H_\mathbf{cluster}$ defined by
\begin{equation}
E = E_\mathbf{s} \otimes \bigotimes_{\mathbf{m} \neq \mathbf{s}}
I_{H_\mathbf{m}}
\end{equation}
where $I_{H_\mathbf{m}}$ is the identity operator acting on $H_{\mathbf{m}}$.
The associated projective measurement on the cluster Hilbert space
$H_\mathbf{cluster}$ is the completely positive map on density states\footnote{We
note that the message of Footnote \ref{reminder} applies to this equation.}
\begin{equation}
P_E\cdot\rho = E \rho E + (1 - E) \rho (1 - E)~.
\end{equation}
The cluster scheme is illustrated in Figure \ref{cluster-configuration-figure}.

%
A general class of cluster-type configurations associated to graphs has been introduced
in the literature~(\cite{raussendorf-browne-briegel-graph-states},
\cite{benjamin-eisert-stace-2005}). Given a graph
$\mathcal{G}=(\mathbf{nodes}, \mathbf{edges})$, the lattice sites are
the elements of $\mathbf{nodes}$.  Each lattice site $a \in
\mathbf{nodes}$ has associated with it a two-dimensional Hilbert space
$H_a$ and an ``entanglement'' projection operator $F$ that acts on the
Hilbert space
\begin{equation}
H_\mathbf{graph} = \bigotimes_{a \in  \mathbf{nodes}} H_a
\end{equation}
as follows:\footnote{Such operators are considered
from the more general point of view as partial
isometries below in Definition \ref{isometric_def} and in the discussion preceding
it.} To each $(a,b) \in \mathbf{edges}$, define the projector
of the form
\begin{equation}
\bigotimes_{n \in \mathbf{nodes} \setminus \{a,b\}}I_{H_n} \otimes
F_{(a,b)}.
\end{equation}
defined in terms of the Pauli matrices by
\begin{equation}
F_{(a,b)} = \frac{1}{2} \biggl\{I + \sigma^{(a)}_z+\sigma^{(b)}_z -
\sigma^{(a)}_z \otimes \sigma^{(b)}_z\biggr\}.
\end{equation}
Then define 
\begin{equation}
F = \prod_{(a,b) \in \mathbf{edges}} F_{(a,b)}.
\end{equation}
$F$ is a projection, since all the projectors $F_{a,b}$ pairwise
commute.  
%

\subsubsection{The Cluster Measurement Scheme}

In the cluster-based approach as explained
in~\cite{raussendorf-briegel-2002}, a computation is realized by a
sequence of projective measurements performed on different sites
$\mathbf{s} \in L$.  The projective measurement that is to be
performed at each step of the computation is determined by a scheme
that specifies at which site to make the measurement and what
observable to measure at that site. These two choices depend on the
outcome of the preceding measurements. In order to specify this
process, a discrete time ``flow'' between the sites is also given.
This flow determines the sequencing of sites to measure. It is
important to note however that the specific projective measurement
taken at each site depends on the outcome of the measurement taken at
the preceding site.

\begin{figure}
\input{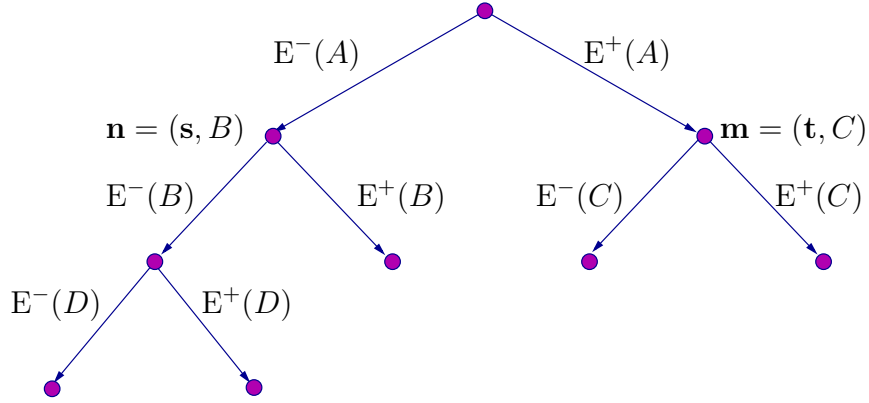}
\caption{Example scheme of a cluster computation}
\label{cluster-scheme-figure}
\end{figure}

At the level of specification, we can also consider the cluster measurement
scheme as given by a multi-rooted tree $\mathbb{T}$.  In
Figure~\ref{cluster-scheme-figure} we illustrate such a tree which
because of spatial limitations is singly rooted. The tree consists of
nodes and directed branches. Each node on the tree $\mathbb{T}$
corresponds to a pair $(\mathbf{s}, A)$ where $\mathbf{s} \in L$ is a
cluster site and $A$ is a two-level observable on $H_\mathbf{s}$.  We
will refer to $\mathbf{s}$ as the cluster site corresponding to the
tree node $(\mathbf{s}, A)$.  Each node $\ell = (\mathbf{s}, A)$ of
the tree has two outgoing branches corresponding to the two possible
outcomes of the measurement of $A$. These two branches correspond to
the spectral projections of $A$, which we denote $\opr{E}^+(A)$,
$\opr{E}^-(A)$.

It is important to note that many different nodes of $\mathbb{T}$ may
correspond to the same cluster site, that is two distinct computation
sequences may take us to the same node but at which different
measurements are taken.  In fact, in the usual cluster approach all
computation sequences traverse the exact same cluster nodes.  This
means that at each horizontal level of the tree $\mathbb{T}$, the
nodes all have the same cluster site.  In the general scheme outlined
above, no such restriction exists.  Thus we may consider schemes in
which not only the subsequent measurement depends on previous
outcomes, but in which the site at which the measurement is taken also
dependent on previous outcomes.

For clusters that are not arranged in a 1 dimensional array, we can
avoid the use of multiply rooted trees if we allow cluster systems
that are not restricted to two-level systems.  This means that the
Hilbert space $H_\mathbf{s}$ corresponding to a site $\mathbf{s} \in
L$ can have arbitrary dimension.  In that case, the corresponding node
observables are allowed to have more than two outcomes and in
particular, the specification tree may not be binary.

A complete path (that is one which originates at the root node and
terminates at a leaf node) of $\mathbb{T}$ is specified by a sequence
of measurement outcomes, where the observable measured is associated
to a node of the tree. Mathematically, each measurement outcome is
expressed by one of the spectral projections associated to the
measurement and graphically represented by an edge of the tree.
A complete measurement outcome associated to a path is a sequence of
projections, where each projection corresponds to a traversal of an
edge of the tree as follows:
\begin{equation}
\ell_0 \stackrel{\opr{E}^\pm_0}{\longrightarrow} \ell_1
\stackrel{\opr{E}^\pm_1}{\longrightarrow} \ell_2
\stackrel{\opr{E}^\pm_2}{\longrightarrow} \ell_3
\stackrel{\opr{E}^\pm_3}{\longrightarrow} \cdots
\stackrel{\opr{E}^\pm_k}{\longrightarrow} \ell_{k+1}.
\end{equation}

The projective measurement associated to the cluster consists of the
projective measurements on sites following the branches of the cluster
scheme tree.  We can explicitly write this down:
\begin{thm}
The projective measurement on a cluster is of the form
\begin{equation}
P_{\mathbb{T}}\cdot\rho = \sum_{\mathcal{P} \in \opr{Path{\mathbb{T}}}}
\biggl(\prod_{\tau \in \mathcal{P}} E_\tau \biggr) \ \rho
\ \biggl(\prod_{\tau \in \mathcal{P}} E_\tau\biggr)
\end{equation}
where $E_\tau$ is a projector on the cluster site corresponding to the
edge of the cluster scheme tree.  Note that for
each $\mathcal{P} \in \opr{Path}(\mathbb{T})$, all the
$E_\tau$ with $\tau \in \mathcal{P}$ commute.
\end{thm}

\subsubsection{Quantum Components in the graph state-based paradigm} 

We now show that the graph state-based paradigm of quantum computation is properly
described by the QCC. The paradigm has been described in the
literature as employing an entangled substrate upon which a series of
conditional projective measurements is performed.  The projective
measurements are used to carry out a computational algorithm, but can
also be used to read input from a macroscopically observable input
register or output to a macroscopically observable output register.
The entangled substrate corresponds to some particular Hilbert
subspace of entangled vectors which may be characterized in various
ways, for instance as the range of an entangling operation or as the
solutions to some eigenvalue equations.

Correspondingly, we should expect that the mathematical formulation of
the graph model of a quantum component also consist of two parts
(We assume as given the Hilbert space $H_{\mathbf{graph}}$):
\begin{enumerate}
\item An initial ``entanglement producing'' operation of some kind.
%
%
\item The projective measurement on $H_{\mathbf{graph}}$
corresponding to the graph scheme.
\end{enumerate}
In our approach, we have already discussed how to formalize the step
(2) above.  However, there are several choices for the entanglement
generating step (1).  In the examples discussed in the literature,
these are given by a self-adjoint projection operator, but it is
natural from our viewpoint to consider more generally partial
isometries $V: H \rightarrow H$.
\begin{dfn}\label{isometric_def}
A {\em graph state-based quantum component} $\mathcal{C}$ is a pair $(V,
\mathbb{T})$ given by a cluster scheme $\mathbb{T}$ and an
``entanglement'' partial isometry $V$ on $H_\mathrm{graph}$.
%
%
The completely positive map associated to $\mathcal{C}$ is defined by
\begin{equation}\label{clust_comp}
Q_\mathcal{C} \cdot\rho = \sum_{\mathcal{P} \in \opr{Path}(\mathbb{T})}
E_\mathcal{P} V \rho V^\dag E_\mathcal{P}
\end{equation}
\end{dfn}
Thus, we see that, just as for the circuit-based paradigm and the adiabatic quantum
computing paradigm, the QCC provides an over-arching framework in which to formulate
the graph-state based paradigm of quantum computing. Moreover, in this section
we have generalized the definition of graph state- (and cluster state-) based
quantum computers that has previously appeared in the literature. Our generalization
consists of two features: (1) our definition allows for {\em arbitrarily
different} measurements to
be carried out at different nodes at the same level of the multi-rooted tree $\mathbb{T}$,
and (2) our definition replaces the use of a self-adjoint projection
operator as an entanglement generator with the more general notion of a partial isometry
(which includes self-adjoint projections as a special case).

\section{Error Thresholds and Fault Tolerance}\label{et_ft}

\subsection{Introduction}

In \S\ref{KLP_reduction} we showed that the recently discovered approach
to error correction known as ``operator quantum error correction" is in fact
a special case of the general QCC formulation. 
Given the general applicability of the QCC to all quantum computing paradigms
({\em cf} \S\ref{unification}), as well as to all
techniques for protection against errors (including
quantum error correction, decoherence-free subspaces and noiseless subsystems),
the QCC thus provides a unified framework for a fully general analysis of
fault tolerance in quantum computing. We
refer to this as operator quantum fault tolerance (OQFT).

As an example of OQFT, in this section we describe the application
of the QCC to the analysis of error thresholds for fault tolerance in the circuit
paradigm.
To make contact with previous research, we begin by discussing this subject from 
the perspective of the well established 
method based on the analysis of error
probabilities~\cite{shor_ft_96}, \cite{aharonov_ben_or_96}, \cite{kitaev1997}, \cite
{klz_97}, \cite{preskill_ft_97}, \cite{agp_ft_05}.
Since the QCC in fact provides the underlying framework in which to study
any issues associated with physical quantum computation, we then reformulate 
the problem in terms of the QCC. This allows us to
relate the results of the two approaches, and to determine the extent to which the
previously utilized approach (based on the method of error probabilities) is in fact
justified based on the insight provided with the QCC.

\subsection{The Method of Error Probabilities}  

In this section we briefly describe the method of error probabilities.
We begin by identifying a quantum operation that we wish to implement and specifying a
circuit 
that (ideally) implements the operation.  
We then identify an error model for the gates in the 
circuit that accounts for the inevitable effects of dissipation and decoherence 
that come into play when the gates are implemented with  
real devices.  By hypothesis, the probability that an error occurs in 
this ``direct" implementation of the 
operation 
is too high for it to be useful as a component in a quantum computer.  

In order 
to render the circuit more fault tolerant, we specify a 
second, more complicated, circuit using quantum error correction.  
The circuit now operates on the set of {\em encoded} qubits 
obtained by encoding the {\em logical} qubits 
of the direct implementation using a QECC.  The gates in the original 
circuit are replaced by collections of gates that operate on the encoded qubits.  
Additional gates are added to carry out the QECC's recovery procedure wherever an error 
is detected.  The error model is then applied to the gates in this new circuit.  
The error correction clearly provides some degree of protection against errors, but
it also necessitates
a larger number of gates, so that there are then more 
opportunities for errors to occur.  In order to determine whether this 
procedure has improved the fault tolerance, we compare the probability 
of an uncorrected error occurring during operation of the QECC version
($\epsilon_f^{\mathrm{QECC}}$) with the 
probability of any error 
occurring in the ``direct" implementation ($\epsilon_f^{\mathrm{direct}}$).
If the QECC error probability is smaller, 
that is, if

\begin{equation}
\label{EQ:stdratio}
\frac {\epsilon_f^{\mathrm{QECC}}}{\epsilon_f^{\mathrm{direct}}} <1 ~,
\end{equation}

\noindent then the procedure has been at least partially successful.  If the error 
probability of the new circuit is still too high, we can reduce the error probability 
further by concatenating the code, that is by  
encoding the qubits used in the QECC version using the same QECC and by redesigning 
the circuit to handle the second level of encoded qubits.
By repeating the process of concatenation we can arrange that the error
probability be made arbitrarily small.

The key point is that (\ref{EQ:stdratio}) can be shown to hold 
only if the failure probabilities of the 
individual gates are less than some threshold value that depends on the relative complexity of 
the encoded {\em vs}. un-encoded versions of the circuit. 
The thresholds appearing in these conditions are known as ``error thresholds." If the 
threshold conditions are satisfied, then the use of concatenated codes will
provide fault tolerant operation to within some 
specified tolerance. The problem of achieving fault tolerant 
quantum computation is reduced to the problem of constructing implementations of the 
primitive gates 
that satisfy the error threshold conditions.  

\subsection{Operator Quantum Fault Tolerance and the QCC}  

We now reconsider the above problem from the perspective of OQFT by making use of
the QCC.
Our goal is to identify a quantum component $P$, that implements 
(approximately) a quantum computation, $U$, that is, we write the quantum
computer condition, $\mathbf{QCC}(P,U,\mathcal{M}_\mathrm{enc},
\mathcal{M}_\mathrm{dec},\alpha)$,

\begin{equation}
\|\mathcal{M}_\mathrm{dec}(P\cdot
(\mathcal{M}_\mathrm{enc}(\rho))) - U \rho U^\dag\|_1 \leq \alpha   
\end{equation}

\noindent for some suitable choice of encoding and decoding
maps, $\mathcal{M}_\mathrm{enc}$ and 
$\mathcal{M}_\mathrm{dec}$. For simplicity of notation we define an operator 

\begin{equation}
\tilde{P} \equiv \mathcal{M}_\mathrm{dec} \cdot P \cdot \mathcal{M}_\mathrm{enc}~,
\end{equation}

\noindent so that the QCC becomes

\begin{equation}
\|\tilde{P} \rho - U \rho U^\dag\|_1 \leq \alpha~.
\end{equation}

We begin by developing a ``zero-th order" implementation 
that does not use a QECC.  In the absence of errors, we assume that the 
implementation faithfully implements the computation $U$:

\begin{equation}\label{EQ:perfectcircuit}
\tilde{P}^{\left( 0 \right)}\rho = 
\mathcal{M}_\mathrm{dec} \left[ V_{\mathrm{circuit}}^{\left( 0 \right)} 
              \left( \mathcal{M}_\mathrm{enc}\rho \right)
     V_{\mathrm{circuit}}^{\left( 0 \right)\dag} \right] = U \rho U^\dag.  
\end{equation}

\noindent With the error model (\ref{EQ:circuiterrormodel}) we have

\begin{equation}
\|\tilde{P}^{\left( 0 \right)}\rho - U \rho U^\dag\| = 
   {\Big \|} \left( 1 - \epsilon_f^{\left( 0 \right)} \right) 
           \mathcal{M}_\mathrm{dec} \left[
              V_{\mathrm{circuit}}^{\left( 0 \right)} 
                   \left( \mathcal{M}_\mathrm{enc}\rho \right) 
              V_{\mathrm{circuit}}^{\left( 0 \right)\dag} \right] +
       \epsilon_f^{\left( 0 \right)} 
              \tilde{Q}_f^{\left( 0 \right)} \rho - U \rho U^\dag {\Big \|}_1 ~,
\end{equation}

\noindent where, for generality, we have subsumed the encoding and decoding maps 
into the definition of $\tilde{Q}_f$.  (The maps are identities for the zero-th order circuit.)  
With (\ref{EQ:perfectcircuit}) the left hand side of the QCC takes the simple form

\begin{equation}
\label{EQ:lhszero}
\|\tilde{P}^{\left( 0 \right)}\rho - U \rho U^\dag\| = \epsilon_f^{\left( 0 \right)} 
   \|  \tilde{Q}_f^{\left( 0 \right)} \rho - U \rho U^\dag  \|_1 ~.
\end{equation}

\noindent Note from \eqref{eps_mu_f} that
the failure probability for this circuit is, to lowest order, 
linear in the error probabilities for the gates which make up the circuit:

\begin{equation}
\label{EQ:epsorderzero}
\epsilon_f^{\left( 0 \right)} \sim \mathcal{O} \left(\epsilon_\mu \right)~,
\end{equation}

\noindent (recall that $\epsilon_\mu$ represents gate error,
{\em cf} \eqref{eps_mu_f})
a fact which, as we shall see, is crucial to the derivation of an 
error threshold.  

We next construct the ``first order" implementation of $U$, which operates in the codespace 
of the QECC.  By the preceding arguments, this implementation will satisfy

\begin{equation}
\label{EQ:lhsone}
\|\tilde{P}^{\left( 1 \right)}\rho - U \rho U^\dag\| = \epsilon_f^{\left( 1 \right)} 
   \|  \tilde{Q}_f^{\left( 1 \right)} \rho - U \rho U^\dag  \|_1 ~.
\end{equation}

Following~\cite{preskill_ft_97}, we
consider the case that errors affect the qubits in the circuit independently 
and that the QECC recovery procedure is sufficient to correct a single error in any one qubit.  
In that case, the 
encoded circuit will exhibit an unrecoverable error only if two or more single qubit errors 
occur.  In this case, the error probability will be quadratic in $\epsilon_\mu$ to leading 
order:  

\begin{equation}
\label{EQ:epsorderone}
\epsilon_f^{\left( 1 \right)} \sim \mathcal{O} \left(\epsilon_\mu \right)^2 ~.
\end{equation}

\noindent  At this point, the QECC has not eliminated all errors, but has resulted in a 
circuit with error probabilities that are quadratic, rather than linear, in the error 
probabilities of the primitive operations.  

We have now introduced two implementations of the computation.  If either implementation 
satisfies the QCC, then 
there is no need to continue.  If the implementations do not satisfy the QCC, then it is meaningful 
to ask whether this can be achieved by concatenating the code.  In order to answer this question, 
we begin by asking another, related question: has the QECC improved the fault tolerance of the 
implementation relative to the QCC?  In other words, we wish to know 
under what conditions it is true that for all $\rho$

\begin{equation}
\label{EQ:levelonecond}
\|\tilde{P}^{\left( 1 \right)}\rho - U \rho U^\dag\|_1 <
 \|\tilde{P}^{\left( 0 \right)}\rho - U \rho U^\dag\|_1 ~,
\end{equation}

\noindent or alternatively\footnote{Relation \eqref{sup_bound} is
actually a stronger condition
than \eqref{EQ:levelonecond} for infinite-dimensional vector spaces.}

\begin{equation}\label{sup_bound}
\opr{sup}_\rho \frac {\|\tilde{P}^{\left( 1 \right)}\rho - U \rho U^\dag\|_1}
 {\|\tilde{P}^{\left( 0 \right)}\rho - U \rho U^\dag\|_1} < 1~.
\end{equation}

\noindent Using (\ref{EQ:lhszero}) and (\ref{EQ:lhsone}), this becomes

\begin{equation}
\label{EQ:epsnormratio}
\opr{sup}_\rho \frac {\epsilon_f^{\left( 1 \right)}}{\epsilon_f^{\left( 0 \right)}}
   \cdot \frac {\|  Q_f^{\left( 1 \right)} \rho - U \rho U^\dag  \|_1}
               {\|  Q_f^{\left( 0 \right)} \rho - U \rho U^\dag  \|_1} < 1~.
\end{equation}

\noindent The above equation represents the extension of \eqref{EQ:stdratio} to OQFT
obtained by using the general framework provided by the QCC.
Clearly it does not reduce to a simple ratio of error probabilities, as in 
(\ref{EQ:stdratio}) above.  The reason for this is that this formulation of the error 
threshold problem based on the QCC takes into account not only error
probabilities, $\epsilon_f$, but 
also expressions involving the norms of operators that characterize the ``strength" 
of the errors.  To see this, note that we began by assuming an error model represented by 
the operation $\tilde{Q}_f^{\left( 0 \right)}$ 
for the zero-th order implementation.  The error model in the 
encoded implementation is represented, in general, by a different operation, 
$\tilde{Q}_f^{\left( 1 \right)}$.  
There is no reason to suppose that these error models are equally effective in perturbing the 
computation.  This is reflected in the difference in the norms:

\begin{equation}
\|  \tilde{Q}_f^{\left( 1 \right)} \rho - U \rho U^\dag  \|_1 \neq
\|  \tilde{Q}_f^{\left( 0 \right)} \rho - U \rho U^\dag  \|_1 ~.
\end{equation}

To further emphasize this point, we obtain the 
above result (\ref{EQ:stdratio}) based on the assumption that the error models 
are ``commensurate" in 
the sense that

\begin{equation}
\label{EQ:strengthapprox1}
\|  \tilde{Q}_f^{\left( 1 \right)} \rho - U \rho U^\dag  \|_1 \approx
\|  \tilde{Q}_f^{\left( 0 \right)} \rho - U \rho U^\dag  \|_1 ~.
\end{equation}

\noindent We shall shortly return to the question of whether this is a good approximation.    
With this assumption, the condition (\ref{EQ:epsnormratio}) then becomes

\begin{equation}
\label{EQ:epsratio10}
\frac {\epsilon_f^{\left( 1 \right)}}{\epsilon_f^{\left( 0 \right)}} \lesssim 1~,
\end{equation}
 
\noindent which is identical to the result (\ref{EQ:stdratio}) obtained by the 
method of error probabilities.
We obtain the form of the error threshold by noting 
from (\ref{EQ:epsorderzero}) and 
(\ref{EQ:epsorderone})
that the numerator and 
denominator of (\ref{EQ:epsratio10}) are, to lowest order, quadratic and linear,
respectively, in $\epsilon_\mu$, so that
\begin{equation}
{\epsilon_f^{(1)}\over\epsilon_f^{(0)}} =
{\sum_{\mu\nu}{A_{\mu\nu}\epsilon_\mu\epsilon_\nu + \cdots}\over
{\sum_\mu B_\mu\epsilon_\mu +\cdots}}~.
\end{equation}
At this point, it is straightforward to obtain a threshold if we set
the $\epsilon_\mu$ equal to each other, and take $\epsilon\equiv\epsilon_\mu$.
Then one obtains the error
threshold comparable to those obtained
in \cite{preskill_ft_97}, \cite{gottesman_ft_97}, \cite{zalka_ft_96}, as
\begin{equation}
\epsilon\lesssim{{\sum_\mu B_\mu}\over{\sum_{\mu\nu} A_{\mu\nu}}}~.
\end{equation}
The desired behavior of 
the concatenated QECC described above follows if we successively apply the 
approximation (\ref{EQ:strengthapprox1}) 
at each level of concatenation:

\begin{equation}
\label{EQ:strengthapprox}
\|  \tilde{Q}_f^{\left( i+1 \right)} \rho - U \rho U^\dag  \|_1 \approx
\|  \tilde{Q}_f^{\left( i \right)} \rho - U \rho U^\dag  \|_1 ~.
\end{equation}

We note that Aliferis, Gottesman and Preskill 
\cite{agp_ft_05} also relate the ratio of error probabilities for successive 
levels of concatenation to an overall measure of the ``accuracy" of the quantum computation.  
Their approach differs from the one described here in three important ways: 

\noindent (Contrast 1) 
Accuracy in \cite{agp_ft_05} is defined in terms of the 
probabilities $p_i$ of the outcomes $i$ of measurements 
on the final output state for an ideal, as compared with a noisy, circuit: 
\begin{equation}
\sum_i |p_i^\mathrm{noisy} - p_i^\mathrm{ideal}|~.  
\end{equation}
In contrast, we define the accuracy of the implementation by the QCC.  

\noindent (Contrast 2) The threshold proofs in \cite{agp_ft_05} rely on proofs that the 
implementations at each level of concatenation are conditionally correct 
relative to the noise model.  
In this case the ratio of error probabilities $\epsilon^{(i+1)} / \epsilon^{(i)}$ 
is automatically the quantity of interest in comparing performance at each level of 
concatenation.  Here, in contrast, the QCC provides the figure of merit at each level
of concatenation, 
and the dependence on error probabilities is inferred.  

\noindent (Contrast 3) As a consequence of the two preceding
points, \cite{agp_ft_05} makes contact with the notion of accuracy only 
at the highest level of concatenation, at which the entire quantum component may be 
viewed as a black box. Here, the QCC is applied systematically at each level 
of concatenation.

At this point we have shown that we can obtain the standard form of the
error threshold result 
from the QCC by introducing the assumption (\ref{EQ:strengthapprox}) on the 
relative strengths of the error models appropriate to successive levels of concatenation 
of the QECC.  We now investigate the validity of the approximation.  
We begin by making some reasonable assumptions about the error operators and the 
initial state of the computer.  We note that the operators 
$\tilde{Q}_f^{\left( i \right)}$ 
are trace preserving, and thus describe the evolution of the quantum component if a 
failure has in fact occurred, as can be seen by
setting $\epsilon_f = 1$ in \eqref{EQ:circuiterrormodel}.
It is then reasonable to expect that the state resulting from its operation on $\rho$ 
will be ``close" to the maximum entropy state\footnote{
Note that this is not a good assumption for errors that operate locally on only one qubit 
in a larger set of 
qubits, leaving the others unaffected.  This important case is a topic for further study.
} in the sense that, for small $\delta_p$,

\begin{equation}\label{EQ:closetomax}
\| \tilde{Q}_f^{\left( i \right)}\rho - \rho_I \|_p < \delta_p~,
\end{equation}

\noindent where $\rho_I$ is the maximum entropy state, and the value of $p$ identifies
the Schatten $p$-norm associated to the corresponding Schatten $p$-class
({\em cf} \eqref{p-norm}).  

\noindent On the other hand, it is normally the case that the input state for the quantum 
computation is a pure state, and thus so is the state $U \rho U^\dag$.  In this case, it 
is straightforward to show that

\begin{equation}
\| \rho_I - U \rho U^\dag \|_1 = 2 - \frac{2}{N}
\end{equation}

\noindent and 

\begin{equation}
\| \rho_I - U \rho U^\dag \|_\infty = 1 - \frac{1}{N}~,
\end{equation}

\noindent where $N$ is the dimension of $H_{\mathrm{logical}}$.  Since 

\begin{equation}
\| \tilde{Q}_f^{\left( i \right)}\rho - U \rho U^\dag \| = 
\| \left[ \tilde{Q}_f^{\left( i \right)}\rho - \rho_I \right] + 
   \left[ \rho_I - U \rho U^\dag \right] \|~,
\end{equation}

\noindent we have, by triangle inequalities, 

\begin{equation}
2 - \frac{2}{N} - \delta_1 \leq 
\| \tilde{Q}_f^{\left( i \right)}\rho - U \rho U^\dag \|_1 \leq
2 - \frac{2}{N} + \delta_1
\end{equation}

\noindent or

\begin{equation}
1 - \frac{1}{N} - \delta_\infty \leq 
\| \tilde{Q}_f^{\left( i \right)}\rho - U \rho U^\dag \|_\infty \leq
1 - \frac{1}{N} + \delta_\infty
\end{equation}

\noindent where we have assumed implicitly that $N \gg 1$ and $\delta_p \ll 1$.  
Since these expressions hold for any value of $i$, (\ref{EQ:strengthapprox}) 
is a reasonable approximation with either choice of norm under the 
conditions that 

\begin{enumerate}
\item (\ref{EQ:closetomax}) holds for some $\delta_p \ll 1$, and 
\item the initial state of the computation is a pure state. 
\end{enumerate}

\section{Conclusions}

In this paper we have presented a fundamental, unifying framework for describing
physically-realizable quantum computing machines. This is concisely stated in the form
of the Quantum Computer Condition (QCC), an inequality that
incorporates a complete specification of the full dissipative, decohering
dynamics of the actual,
practical device used as the quantum computing machine, a specification of
the ideally-defined quantum
computation intended to be performed by the machine, and a quantitative criterion
for the accuracy with which the computation must be executed. 

With the QCC we prove the fundamental Encoding No-Go Theorem that identifies the amount of 
damping (including dissipative and decohering effects) for which physically-realizable
fault-tolerant quantum computing is not possible.
We provide a rigorous definition of damping, and
explicitly calculate a {\em universal} critical damping
value for fault-tolerant quantum computation.
This theorem can be used in principle
to solve practical problems involving quantum computer design. 

In this paper we have also presented an existence proof for fundamental solutions to useful
classes of {\em time-dependent} generalizations of the Lindblad equation. This can provide
a useful tool in analyzing a wide variety of open quantum mechanical systems.

We have demonstrated that the entire formalism of operator quantum error correction
(OQEC) can be obtained from the QCC as a special case.
By allowing for the possibility of residual errors, the general formalism of the QCC
enables us to generalize OQEC to ``operator quantum fault tolerance" (OQFT).
Since we have demonstrated that
OQEC is in fact a particular reduction of the QCC, and since standard
quantum error correction (QECC), decoherence-free subspaces (DFS) and noiseless
subsystems are all special cases of OQEC, we have discovered that QCC applies in general
across {\em all} these approaches.

As an initial application of the OQFT concept, we
have begun the exploration of the application of QCC to the problem of establishing
thresholds for fault-tolerant quantum computation by showing that the standard
approaches to this problem can be motivated within the framework of the QCC.

Research in quantum information science has resulted in the discovery of seemingly different
{\em paradigms} for quantum computation, including the circuit-based paradigm,
graph state-based paradigm and adiabatic quantum computing paradigm. In this
paper we have explicitly demonstrated
that these paradigms are not in fact distinct at a fundamental level,
but are all describable within the unifying framework provided by the QCC.
In the particular case of the graph state-based paradigm (which includes cluster
state-based models), we not only show that the
paradigm is a manifestation of the unifying picture provided by the QCC,
but also introduce a definition of graph state-based quantum computers that
generalizes the graph state models previously defined in the literature.

Future work motivated by our results should include applications of the Encoding
No-Go Theorem to diverse problems pertaining to practical quantum computer design
and implementation.
It would also be of interest to further explore the physics of the operator
quantum fault tolerance (OQFT) generalization of OQEC presented in this paper.
Specific work along theses lines should include
further application of the QCC to obtaining error thresholds for fault-tolerant
implementations of quantum computers.
It would also be fruitful to explore applications of the quantum computer
condition
to situations in which quantum process tomography techniques are used to
experimentally characterize the quantum component described by the completely positive
map that appears in the QCC.\\

\noindent{\em Acknowledgements}

\noindent We wish to thank Anthony Donadio and Yaakov Weinstein for helpful comments.
This work was supported by MITRE under the MITRE Technology Program.

\appendix

\section{Banach Spaces of Operators} \label{banach-spaces} A linear
map $T$ on a Banach space is a {\em contraction} iff its norm is $\leq
1$.

Let $H$ be a separable Hilbert space.  $\mathbf{L}(H)$ denotes the
space of bounded operators on $H$ with the operator norm,
$\mathbf{K}(H)$ denotes the normed closed subspace of compact
operators of $\mathbf{L}(H)$.  In case $H$ is finite dimensional,
these spaces are identical.  Assume now $H$ is infinite dimensional;
we consider other Banach spaces of compact operators defined by
eigenvalue decay conditions and whose norms reflect the rate of decay
of the eigenvalues.  Specifically, let $T \in \mathbf{K}(H)$, then
$|T| = \sqrt{T^\dag T}$ is a non-negative compact operator and so by
the spectral theorem, has a complete set of eigenvectors with
eigenvalues that can be ordered in a sequence
\begin{equation}
s_0(T) \geq s_1(T) \geq \cdots \geq s_n(T) \geq 0
\end{equation}
which converges to $0$.  The following properties are well-known
(see~\cite{gohberg-krein}; also~\cite{connes} on which this discussion
is based). 

$\mathbf{T}(H)$ is the Banach space of trace-class
operators $T$ on $H$ with the norm
\begin{equation}
\|T\|_1 = \sum_{k=0}^\infty | s_k(T) |
\end{equation}
More generally, the Schatten $p$-class $\mathbf{T}_p(H)$ is defined by
the condition
\begin{equation}\label{p-norm}
\|T\|_p = \bigg\{\sum_{k=0}^\infty s_k(T)^p \biggl\}^\frac{1}{p} < \infty
\end{equation}
with the norm given by $\| \cdot\|_p$.
The operator norm for any $T\in \mathbf{L}(H)$, denoted by $\|
T\|_\infty$, is defined as the supremum of $\| T x \|$ for $x \in H$
of norm $\leq 1$.  If $T \in \mathbf{K}(H)$, the operator norm is also
the supremum of the eigenvalues of $|T|$.

\section{Completely Positive Maps and Fundamental Solutions} 
\label{mathematical-preliminaries-section}

In the following, $H$ denotes a complex Hilbert space. We will
consider various Banach spaces of bounded operators on $H$: these are
discussed in Appendix~\ref{banach-spaces} above. 

We will consider completely positive maps on both $\mathbf{L}(H)$ on
the Schatten classes $\mathbf{T}_p(H)$ and especially on the
trace-class operators $\mathbf{T}(H) = \mathbf{T}_1(H)$.

The basic fact about completely positive maps we use is the {\em Kraus
representation}.  
\begin{prop}
A $P:\mathbf{T}(H) \rightarrow \mathbf{T}(H)$ is a completely positive contraction
if and only if it is of the form
\begin{equation}\label{KrausFormCPMap}
P(\rho)=\sum_{i \in I} X_i \rho X_i^\dag
\end{equation}
where $X_i \in \mathbf{L}(H)$ with
\begin{equation}
\sum_i X_i^\dag X_i \leq 1
\end{equation}
\end{prop}
We will consider only {\em trace preserving} completely positive maps $P$,
that is which satisfy
\begin{equation}\label{trace-norem}
\opr{tr}(P(\rho)) = \opr{tr}(\rho)
\end{equation}

\begin{prop} Suppose $P$ is a completely positive map given by the
Krauss form~\eqref{KrausFormCPMap}.
\begin{enumerate}
\item A necessary and sufficient condition $P$ be trace-preserving is that
\begin{equation}\label{tr-preserving-cond}
\sum_{i \in I} X_i^\dag  X_i =1.
\end{equation}
\item  A necessary and sufficient condition that $P$ be bounded in the operator norm is that
\begin{equation}\label{norm-preserving-cond}
\sum_{i \in I} X_i  X_i^\dag \in \mathbf{L}(H).
\end{equation}
\end{enumerate}
\end{prop}
Note that if $T$ is a completely positive trace-preserving and
operator norm continuous map, then by interpolation $T$ is also norm
continuous on the Schatten $p$-classes.

\begin{prop}
If $P$ is a completely positive trace-preserving map, then the adjoint of
$P$ on $\mathbf{L}(H)$ defined by
\begin{equation}\label{definition-of-dual}
\opr{tr}(P^{\mathrm{t}}(T) \rho) = \opr{tr}(T P(\rho))
\end{equation}
is a unit preserving completely positive map
$\mathbf{L}(H) \rightarrow \mathbf{L}(H)$.  Its Kraus representation is
\begin{equation}
P^{\mathrm{t}}(T)=\sum_{i \in I} X_i^\dag T  X_i
\end{equation}
\end{prop}
Note that the completely positive unit preserving
maps $\mathbf{L}(H) \rightarrow \mathbf{L}(H)$ which are adjoints of
completely positive trace-preserving maps on $\mathbf{T}(H)
\rightarrow \mathbf{T}(H)$ can be characterized precisely as those
which are continuous mappings $\mathbf{L}(H) \rightarrow
\mathbf{L}(H)$, where $\mathbf{L}(H)$ has the ultraweak topology.

\subsection{Completely Positive Semigroups on $\mathbf{T}(H)$}

We need to first establish that each generalized Lindblad-type
operator $A(t)$ ({\em cf} eqs.\eqref{generic-equation-of-evolution} and \eqref
{LindbladOperator}) generates
a semigroup of completely positive contractions on $\mathbf{T}(H)$
{\em with respect to the trace-class norm $\| \cdot \|_1$}.  We will
also consider boundedness properties relative to the operator norm $\|
\cdot \|_\infty$, although in general the corresponding semigroups may
not be contraction semigroups in this norm.

We begin with a general result characterizing infinitesimal generators
of strongly continuous positive (or completely positive) semigroups on
the Banach space $\mathbf{T}(H)$.  Recall that a one-parameter
semigroup $\{T_t\}_{t \geq 0}$ on a Banach space $E$ is said to be of
class $C_0$ iff for every $x \in E$, $\lim_{t \rightarrow s} T_t(x) =
T_s(x)$.  If $\{T_t\}_{t \geq 0}$ is a $C_0$-semigroup, then there are
positive constants $M$ and $\beta$ such that
\begin{equation}\label{exponential-condition}
\|T_t\| \leq M e^{t \beta}.
\end{equation}
Moreover, 
\begin{equation}
A x = \lim_{h \rightarrow 0} h^{-1} (T_h x - x)
\end{equation}
is a densely-defined operator called the {\em infinitesimal generator} of
$\{T_t\}_{t \geq 0}$

\begin{thm}
Suppose $A$ is a densely-defined operator on $\mathbf{T}(H)$ which
generates a contractive semigroup $\{T_t\}_{t \geq 0}$ on
$\mathbf{T}(H)$.  A necessary and sufficient condition the operators
$\{T_t\}_{t > 0}$ be positive (respectively completely positive) is
that for each $\lambda > 0$
\begin{equation}
\opr{R}(\lambda,A)= (\lambda I - A)^{-1}
\end{equation}
(which is defined by the Hille-Yosida Theorem) be positive
(respectively completely positive).  The operators $T_t$ are
trace-preserving iff in addition for all $\lambda > 0$,
\begin{equation} \label{trace-dilation-property}
\opr{tr}(\opr{R}(\lambda,A) \rho) = \lambda^{-1} \opr{tr}(\rho)
\end{equation}
for every $\rho \in \mathbf{T}(H)$. 

The family of operators $\{T_t\}_{t > 0}$ extends to a $C_0$-semigroup
on $\mathbf{L}(H)$
iff there are constants $M'$ and $\beta'$ such that for all $\lambda
>\beta'$ and for all $\rho \in \mathbf{T}(H)$ and non-negative
integers $m$:
\begin{equation}
\|\opr{R}(\lambda,A)^m \rho\|_\infty \leq M'(\lambda - \beta')^{-m}
\|\rho\|_\infty.
\end{equation}
In this case, we have the explicit bound
\begin{equation}
\|T_t\|_\infty \leq M' e^{t \beta'} \quad \forall t>0.
\end{equation}
\end{thm}
\begin{rem}
It suffices that the property~\eqref{trace-dilation-property} hold for
density states $\rho$.
\end{rem}

\begin{proof}
To avoid duplication, we refer only to the assertions for complete
positivity.  By the general Hille-Yosida theory, if $A$ is an
infinitesimal generator of a contractive semigroup on $\mathbf{T}(H)$,
the resolvents
\begin{equation}
\opr{R}(\lambda,A) = (\lambda - A)^{-1}
\end{equation}
are defined for all $\lambda > 0$ and by Theorem 3.1.3 of~\cite{tanabe},
\begin{equation}\label{Laplace-transform}
\opr{R}(\lambda,A) = \int_0^\infty e^{-\lambda t} T_t dt.
\end{equation}
{\em i.e}., the resolvent is the Laplace transform of $\{T_t\}_{t \geq 0}$.
In particular, if $T_t$ consist of completely positive operators, then
the resolvent operators are all completely positive.  

Conversely, let
\begin{equation}
A_\lambda = \lambda (\lambda  \opr{R}(\lambda,A) - I)
\end{equation}
Then for each $t \geq 0$,
\begin{equation}
\exp (t A_\lambda) = e^{-\lambda t} \exp \big(t \lambda^2 \opr{R}(\lambda,A)\big)
\end{equation}
which is clearly completely positive and it is known that for each $t
\geq 0$,
\begin{equation} \label{semigroup-approximation-equation}
T_t = \lim_{\lambda \rightarrow \infty} \exp (t A_\lambda)
\end{equation}
in the strong operator topology.  Thus $T_t$ is completely positive.

To deal with the trace preservation properties of $T_t$, note that if
$S$ is a bounded operator on $\mathbf{T}(H)$ for which
\begin{equation}
\opr{tr}(S \rho) = \alpha \opr{tr}(\rho)
\end{equation}
then 
\begin{equation}
\opr{tr}(e^S \rho) = \sum_{k=0}^\infty \opr{tr}\bigg(\frac{S^k}{k!} \rho\bigg)=
\sum_{k=0}^\infty\frac{\alpha^k}{k!} \opr{tr}(\rho) = e^{\alpha} \opr{tr}(\rho).
\end{equation}
Thus, 
\begin{equation}
\opr{tr}(\exp t A_\lambda \rho) = e^{-\lambda t} e^{t \lambda^2 \ 1
/\lambda} \opr{tr}(\rho) = \opr{tr}(\rho).
\end{equation}
By~\eqref{semigroup-approximation-equation}, it follows that $T_t$ is also
trace preserving.  
Conversely, if $T_t$ is trace preserving,
\begin{equation}
\opr{tr}(\opr{R}(\lambda,A) \rho) = \int_0^\infty e^{-\lambda t}
\opr{tr}(T_t \rho) dt = \int_0^\infty e^{-\lambda  t}\opr{tr}(\rho) dt = \lambda^{-1}\opr{tr}(\rho).
\end{equation}
\end{proof}
\subsubsection{Examples of Completely Positive Semigroups}

Our analysis of the generalized Lindblad equation will reduce to an analysis of
the solution in two important cases:

\paragraph{(Case 1) Unitary Evolution} 

In particular, if $\mathcal{H}$ is a self-adjoint operator on a
Hilbert space $H$, then the family of completely positive mappings
\begin{equation}
P^\mathcal{H}_t(\rho) = e^{i t \mathcal{H}} \rho e^{- i t \mathcal{H}}
\end{equation}
is a one-parameter group of completely positive maps. Its generator on
the trace-class operators is formally given by the operator
\begin{equation} \label{formal-infinitesimal-generator}
\rho \mapsto i [\mathcal{H},\rho].
\end{equation}
This expression is only formal, because it is not defined for all
$\rho$. Nevertheless, the infinitesimal generator is densely defined
on the space of trace-class operators and it is an extension
of~\eqref{formal-infinitesimal-generator} for the finite rank
operators on the domain of $\mathcal{H}$.\\

\paragraph{(Case 2) Dissipative Operators} 

Another type of infinitesimal
generator we will consider are operators of the form

\begin{equation}\label{pure-lindblad-form}
\mathcal{L} \rho = \sum_j \bigg[ L_j \rho L_j^\dag  - \frac{1}{2}\big\{L_j^\dag
L_j, \rho\big\}\bigg] \end{equation}
where braces denote the anti-commutator.
\begin{lem}  Suppose
\begin{equation} \label{boundedness-cond-of-lindblad}
\sum_j L_j^\dag L_j  \in \mathbf{L}(H).
\end{equation}
Then the operator given by~\eqref{pure-lindblad-form} is bounded on
$\mathbf{T}(H)$. If in addition
\begin{equation} \label{norm-boundedness-cond-of-lindblad}
\sum_j L_j L_j^\dag  \in \mathbf{L}(H).
\end{equation}
then $\mathcal{L}$ is a bounded operator on $\mathbf{L}(H)$.
\end{lem}
\begin{proof} Let $C$ be the operator norm of $\sum_j L_j^\dag L_j$.
To show the map $\mathcal{L}$ is defined and continuous on $\mathbf{T}(H)$, it
suffices to show
$
\mathcal{L}_0: \rho \mapsto  \sum_j L_j \rho L_j^\dag
$
is defined and continuous on $\mathbf{T}(H)$.  However, if $\rho \geq 0$,
\begin{align}
\opr{tr}(\mathcal{L}_0(\rho)) & = \sum_i \opr{tr}(L_j \rho L_j^\dag) \\
& = \sum_j \opr{tr}(\rho L_j^\dag L_j) \\
& = \sum_j \opr{tr}\biggl(\rho^{1/2} L_j^\dag L_j\rho^{1/2}\biggr) \\
& = \opr{tr}\biggr(\rho^{1/2}  (\sum_j L_j^\dag L_j) \rho^{1/2}\biggr) \leq C \opr{
tr}(\rho) = C \|\rho\|_1
\end{align}
thus for arbitrary self-adjoint $\rho$,
\begin{align}
\|\mathcal{L}_0(\rho)\|_1 & = \|\mathcal{L}_0(\rho^+ - \rho^-)\|_1 \\
& \leq  \|\mathcal{L}_0(\rho^+)\|_1 +  \|\mathcal{L}_0(\rho^-)\|_1 \\
& \leq  C\biggl(\|\rho^+\|_1 + \|\rho^-\|_1\biggr) = C \|\rho\|_1.
\end{align}

If~\eqref{norm-boundedness-cond-of-lindblad}, suppose $0 \leq T \leq
1$:
\begin{equation}
0  \leq \sum_j L_j^\dag T L_j \leq \sum_j  L_j^\dag L_j \leq C 1_H
\end{equation}
Thus, for arbitrary $T$,
\begin{equation}
\|\mathcal{L} T \|_\infty \leq  \|\sum_j L_j L_j^\dag\|_\infty  + C \|T\|_\infty
\end{equation}
\end{proof}

It was established by Lindblad~\cite{lindblad76} (and not too hard to show directly) that
if~\eqref{boundedness-cond-of-lindblad} holds, the
$\mathcal{L}$ generates a {\em uniformly continuous} a completely
positive semigroup (relative to the operator norm on $\mathbf{T}(H)$).
In this case the semigroup is given by 
\begin{equation}
e^{t \mathcal{L}} = \sum_{k=0}^\infty \frac{t^k}{k!}{\mathcal{L}}^k~.
\end{equation}
%

\subsection{Perturbation of Completely Positive Generators}

In order to show that the generalized Lindblad operators given in
equation~\eqref{LindbladOperator} are completely positive generators,
we need to establish a perturbation result analogous to the
Kato-Rellich theorem. 

A linear map $B$ on $\mathbf{T}(H)$ is {\em trace annihilating} iff
\begin{equation}
\opr{tr}(B \rho) = 0
\end{equation}
for all $\rho \in \mathbf{T}(H)$.  For example, an operator of the form~\eqref{pure-lindblad-form}
is easily seen to be trace annihilating.
%

%
Using the Trotter-Kato product formula~(\cite{trotter59}, \cite{chernoff68}), we can show that generators of
completely positive contractive semigroups have a sum which is also a
generator of a contractive semigroup, provided the sum generates a
contractive semigroup.

\begin{prop}
Suppose $B$ is a bounded operator of the
form~\eqref{pure-lindblad-form}.  If $A$ is a generator of a completely
positive semigroup then so is $A+B$. .
\end{prop}
%
\begin{cor}
The generalized Lindblad operators given in equation~\eqref{LindbladOperator}
generate a completelty positive semigroup of contractions on $\mathbf{T}(H)$.
\end{cor}

We now extend the results of Lindblad and Davies to allow for time
varying Hamiltonians by relying on results of Kato.  The need for this
arises since in some circuit-based models the various gates are
implemented by varying the Hamiltonian (see for
instance~\cite{nielsen-chuang} \S7.7.2). We make the assumption that
the dissipative effects are bounded which simplifies the analysis
considerably.

\subsection{Solving the Generalized Lindblad Equation}\label{solve_lindblad}

In some cases it is possible to solve
the generalized Lindblad equation~\cite{LuYangZangChen2003}. However, by solution
we mean an expression for the fundamental solution $P_{t,s}$ as a
limit of product of exponentials. Though this expression will almost
never provide a closed form solution, it will provide enough
information to obtain an estimate of how well a unitary (or partial
isometry) can be implemented by one of the operators $P_{t,s}$.  The
two tools we use are the Trotter-Kato product formula and the explicit
form of the solution of a time-dependent equation as a time ordered
product of exponentials given in the proof of  \S4.2
of~\cite{tanabe}.

A precise formulation of a set of conditions which guarantees the
convergence of the products in the next two theorems is given in
Theorem~\ref{precise-formulation-thm}.  These results comprised
by Theorems \ref{ordered-product-exponentials-thm} and
\ref{ordered-product-exponentials-thm2}
are restatements
of assertions contained in the proofs in \S4.2 of~\cite{tanabe}.

\begin{thm} \label{ordered-product-exponentials-thm} Under suitable
conditions, the fundamental solution $P_{t,s}$
for~\eqref{fundamental-solution-def} is given by
\begin{equation} \label{solution-product-formula} P_{t,s} =
\opr{str-lim}_{\Delta \rightarrow 0} \prod_{k=0}^{n-1}
\exp\bigl((r_{k+1} - r_k) A(r_k)\bigr),
\end{equation}\label{ordered-product-exponentials}
where $s = r_0 < r_1 < \cdots < r_{n-1} < t$ and $\max | r_{k+1} -
r_k| \leq \Delta$.
Each $P_{t,s}$ is completely positive and trace preserving.
\end{thm}

\begin{thm} \label{ordered-product-exponentials-thm2} Under
the same assumptions as the previous
theorem~\ref{ordered-product-exponentials-thm},
\begin{equation}\label{Trotter-Lie-exponentials}
\exp s A(t) = \opr{str-lim}_{n \rightarrow \infty} \exp
\biggl(\frac{s}{n} \mathcal{L}(t)\biggr) \ \exp\biggl(\frac{s}{n} \mathcal{H}(t)\biggr)
\end{equation}
\end{thm}

\subsection{Existence of Solutions}\label{existence-and-uniqueness}

We will restrict our attention to bounded time varying perturbations
of a fixed self-adjoint operator acting on $\mathbf{T}(H)$ via a
commutator as in~\eqref{basic-operator} below. The result we state is not
the most general possible, and the early results of Kato~\cite{kato}
suffice for its proof. We follow the treatment in Chapter XIV, \S4
of~\cite{yosida} which is a more readily available reference.

\begin{thm}\label{precise-formulation-thm}
Suppose $\mathcal{H}$ is a self-adjoint operator, $\{B(t)\}_{t \in [0,
\infty[}$, $\{L_j(t)\}_{t \in [0, \infty[}$, $1 \leq j \leq n$ are
families of bounded operators, all of which are continuously norm differentiable as
a functions of $t$, then there is a fundamental solution $P_{t,s}$
for~\eqref{fundamental-solution-def} where
\begin{equation} \label{basic-operator}
A(t) \rho = -i[\mathcal{H}+B(t), \rho] + \sum_{j=1}^n \bigg( L_j(t)
\rho L_j^\dag(t) - \frac{1}{2}\big\{L_j^\dag(t) L_j(t),
\rho\big\}\bigg).
\end{equation}
The solution is a given by a limit of a time-ordered product of
exponentials~\eqref{solution-product-formula}.
\end{thm}
\begin{proof}
There are various technical assumptions for a family $A(t)$ of
operators that need to be checked in order to apply Kato's Theorem.
The first of these is the independence of $\opr{dom} A(t)$ of the
parameter $t$. Under our assumptions
\begin{equation}
A(t) = A + C(t)
\end{equation}
where $C(t): \mathbf{T}(H) \rightarrow \mathbf{T}(H)$ are bounded
operators and $A$ is the infinitesinal generator of a contractive
semigroup.  Indeed, 
\begin{equation}
A\rho  =  -i[\mathcal{H}, \rho] 
\end{equation}
is the infinitesimal generator of a group on $\mathbf{T}(H)$ and 
\begin{equation}
C(t) \rho = -i[B(t), \rho] + \sum_{j=1}^n \bigg( L_j(t)
\rho L_j^\dag(t) - \frac{1}{2}\big\{L_j^\dag(t) L_j(t),
\rho\big\}\bigg).
\end{equation}
is by assumption a bounded operator on $\mathbf{T}(H)$. In particular,
all the operators $A(t)$ have the same domain $\opr{dom}(A)$.

We now address the remaining assumptions in Kato's theorem.  For any
$\lambda >0$, \begin{equation}
\lambda - A(t) = \lambda - A - C(t) = (I +  C(t) \opr{R}(\lambda, A) ) (\lambda - A) 
\end{equation}
For $\lambda$ sufficiently large $\opr{R}(\lambda, A) C(t)$ has norm
$<1$, so the Neumann (geometric) series for inverses
(see~\cite{Dieudonne}, Chapter VIII, \S3) $I + \opr{R}(\lambda, A)
C(t)$ is invertible.  Thus we can write,
\begin{equation}
(\lambda - A(t))^{-1} = \opr{R}(\lambda, A) (I + \opr{R}(\lambda, A) C(t))^{-1} 
\end{equation}
Thus,
\begin{equation}
B(t,s) = (\lambda - A(t))(\lambda - A(s))^{-1} = (I + C(t)
\opr{R}(\lambda, A) ) (I + C(s) \opr{R}(\lambda, A) )^{-1}
\end{equation}
is well defined and by our assumptions $B(t,s)$ is a norm
differentiable function jointly in the variables $t,s$. This implies
the remaining conditions in the hypothesis of Kato's theorem.  It only
remains to observe that presence of the parameter $\lambda$, instead
of $1$ as actually stated in Kato's theorem is immaterial, since
solutions of equations
\begin{equation}
\frac{d}{dt} \rho (t) = A(t) \rho(t)
\end{equation}
are trivially affected by adding a constant scalar to $A(t)$.
\end{proof}

\section{Solving the Ersatz Quantum Computer Condition}\label{alpha_0_solution}

%

Consider the ersatz quantum computer condition (${\mathcal E}$QCC) given
in \eqref{ersatz-quantum-computer-equation}:
\begin{equation} \label{eqn:defining-part}
P\cdot\rho = U \rho U^\dag.
\end{equation}
%
Note that this can be obtained
by setting the encoding and decoding maps to unity, and
setting $\alpha = 0$ in the QCC given in \eqref{encoded_QCC}.
A simple observation shows that
``solving''~\eqref{eqn:defining-part} is completely equivalent to
obtaining the noise-free part of a communication channel.

\begin{thm}\label{krebs-reduction-trick} Suppose $H$ is
finite-dimensional. If $P$ is given by the Kraus
representation~\eqref{KrausFormCPMap},
given a unitary $U$, the set of $\rho \in \mathbf{L}(H)$
satisfying~\eqref{eqn:defining-part} is the $\ast$-subalgebra of
$\mathbf{L}(H)$ given by
\begin{equation}\label{commutant-condition}
\mathfrak{A}_{P,U}= \{\rho \in \mathbf{L}(H): \forall i \in I, \quad
[\rho, U^\dag X_i] = 0 \}
\end{equation}
\end{thm}
\begin{proof}  The solutions of~\eqref{eqn:defining-part} are the
fixed points of the completely positive map $Q$ defined by the equation
\begin{equation}
Q\cdot\rho = \sum_{i \in I} U^\dag X_i \rho X_i^\dag U.
\end{equation}
Now apply~\cite{kribs03}, Theorem 2.1.
\end{proof}

In the above theorem, the assumption $H$ is finite-dimensional is
essential, see~\cite{arias-at-al-2002}. Since $H$ is finite dimensional
$\mathfrak{A}$ is an algebraic direct sum of algebras isomorphic to
full matrix algebras.
\begin{prop}
Let $\{E_\kappa\}_{\kappa \in I}$ be the set of finite-dimensional
minimal central projections of $\mathfrak{A}$.  Then 
\begin{equation}
\mathfrak{A}_\kappa = E_\kappa \mathfrak{A} E_\kappa
\end{equation}
is an algebra of operators on the range $H_\kappa$ of $E_\kappa$
(which is a finite dimensional space). It is isomorphic to a full
matrix-algebra of finite multiplicity.

\end{prop}

\bibliography{refs}
\bibliographystyle{hplain}

\end{document}